\documentclass[a4paper]{article}
\def\wsrmp{0}

\if1\wsrmp

\makeatletter
\let\url\@undefined
\def\@nameundef#1{\expandafter\let\csname #1\endcsname \@undefined}
\@nameundef{equation*}
\@nameundef{endequation*}
\def\@opargbegintheorem#1#2#3{%
   \par\addvspace{6.25pt plus2pt minus.25pt}%
   \def\@tempa{#3}%
   {\bfseries
     \noindent#1 #2\ifx\@tempa\empty\unskip.\else\unskip.\hskip.5em{(#3)}\fi}\hskip1em
   \csname #1bodyfont\endcsname\ignorespaces} 
\renewcommand\appendix{\par
        \refstepcounter{appendix}
        \setcounter{section}{0}%
        \setcounter{lemma}{0}
        \setcounter{theorem}{0}
        \setcounter{definition}{0}
        \setcounter{corollary}{0}
        \setcounter{equation}{0}
        \@addtoreset{equation}{section}
        \def\@seccntformat##1{{\upshape{\appendixname\ \csname the##1\endcsname.}}\hskip .5em}
\renewcommand\thesection{\Alph{section}}
\renewcommand\thesubsection{\Alph{section}.\arabic{subsection}}%
\renewcommand\theequation{\Alph{section}.\arabic{equation}}}%
\makeatother

\fi

\usepackage{amsmath}
\if0\wsrmp
\usepackage[psamsfonts]{amssymb}
\fi
\usepackage{latexsym}
\usepackage{cmmib57}
\usepackage{graphicx}
\usepackage[dvips]{hyperref}

\newcommand{\goth}[1]{\mathfrak{#1}}
\newcommand{\scr}[1]{\mathcal{#1}}
\def\idty{{\leavevmode{\rm 1\ifmmode\mkern -4.8mu\else\kern -.3em\fi
      I}}}
\renewcommand{\Bbb}[1]{\if1#1\idty\else\mathbb{#1}\fi}
\newcommand{\kb}[1]{|#1\rangle\langle#1|}
\newcommand{\ket}[1]{|#1\rangle}
\newcommand{\tr}{\operatorname{tr}}
\newcommand{\Lz}{\operatorname{L}^2}
\newcommand{\SP}{\operatorname{span}}
\newcommand{\diag}{\operatorname{diag}}
\newcommand{\U}{\operatorname{U}}
\newcommand{\GL}{\operatorname{GL}}
\newcommand{\Sym}{\operatorname{S}}
\newcommand{\Id}{\operatorname{Id}}
\newcommand{\PM}{\operatorname{pm}}

\newcommand{\Hz}{\operatorname{H}^2}

\if1\wsrmp

\newtheorem{thm}{Theorem}[section]

\newtheorem{defi}[thm]{Definition}

\newtheorem{prop}[thm]{Proposition}

\newtheorem{lem}[thm]{Lemma}

\newtheorem{cor}[thm]{Corollary}

\else

\newtheorem{thm}{Theorem}[section]
\newtheorem{defi}[thm]{Definition}
\newtheorem{prop}[thm]{Proposition}
\newtheorem{lem}[thm]{Lemma}
\newtheorem{cor}[thm]{Corollary}

\fi

\newenvironment{pf}{\par\noindent\textbf{Proof.\ }}{\hfill $\Box$ \medskip}

\if1\wsrmp

\title{Quantum state estimation and large deviations}
\author{ M. Keyl}
\address{ Istituto Nazionale di Fisica della Materia, Unita' di Pavia,\\
   Dipartimento di Fisica ``A. Volta'', via Bassi 6, I-27100 Pavia, Italy\\
   \email{M.Keyl@tu-bs.de}}

\else 

\title{Quantum state estimation and large deviations}
\author{ M. Keyl\thanks{Electronic Mail: \tt{m.keyl@tu-bs.de}}
  \\[1ex]
  {\small Istituto Nazionale di Fisica della Materia, Unita' di Pavia,}\\
  {\small Dipartimento di Fisica ``A. Volta'', via Bassi 6, I-27100 Pavia, Italy}} 
\date{\today}

\fi

\begin{document}

\maketitle

\if1\wsrmp

\begin{history}
\received{(X.Y.2005)}
\revised{(A.B.2005)}
\end{history}

\fi

\begin{abstract}
  In this paper we propose a method to estimate the density matrix $\rho$ of a $d$-level quantum system by
  measurements on the $N$-fold system in the joint state $\rho^{\otimes N}$. The scheme is based on covariant
  observables and representation theory of unitary groups and it extends previous results concerning pure
  states and the estimation of the spectrum of $\rho$. We show that it is consistent (i.e. the original input
  state $\rho$ is recovered with certainty if $N \to \infty$), analyze its large deviation behavior, and calculate
  explicitly the corresponding rate function which describes the exponential decrease of error probabilities
  in the limit $N \to \infty$. Finally we discuss the question whether the proposed scheme provides the fastest
  possible decay of error probabilities.
  \end{abstract}

\if1\wsrmp



\fi

%%%%%%%%%%%%%%%%%%%%%%%%%%%%%%%%%%%%%%%%%%%%%%%%%%%%%%%%%%%%%%%%%%%%%%%%%%%%%%%%%%%%%%%%%%%%%%%%%%%%
\section{Introduction}
\label{sec:introduction}
%%%%%%%%%%%%%%%%%%%%%%%%%%%%%%%%%%%%%%%%%%%%%%%%%%%%%%%%%%%%%%%%%%%%%%%%%%%%%%%%%%%%%%%%%%%%%%%%%%%%

The density operator $\rho$ of a $d$-level quantum system ($d \in \Bbb{N}$) describes the preparation of the system
in all details relevant to statistical experiments, and the task of quantum state estimation is to determine
$\rho$ by measurements on a (possibly large) number $N$ of systems, which are all prepared according to $\rho$. In
the limit of infinitely many input systems it is of course possible to get exact estimates. If $N$ remains
finite, however, estimation errors are unavoidable. The best we can get (if $N$ is large enough) is an
estimation scheme which produces only small errors or, better to say, which produces large errors only with a
small probability.  

There are several ways to get ``good'' estimation schemes. One possibility is to choose an appropriate figure
of merit which measures the quality of the estimates (e.g. averaged fidelities with respect to the original
density matrix)  and to solve the corresponding optimization problem. If we know a priori that the input state
$\rho$ is pure (but otherwise unknown) this approach is very successful and leads to optimal estimators, which can
be given in closed form for all finite values of $N$;
cf. e.g. \cite{HolBook,MasPop95,BDiVEFMS,DeBuEk98,Hayashi98,LPT98,Bruss99}. In the general case, however
(i.e. if nothing is known about $\rho$) the situation is much more difficult. First of all the result depends
much more on the figure of merit chosen than in the pure state case, and even if we have found an appropriate
quality criterion it is in general very hard to determine the corresponding optimal estimator explicitly for
arbitrary $N$; some results related to this approach can be found in \cite{Vidal99,Fischer00,BBMR04}. 

A way out of this dilemma, is to neglect the quality of the estimates for finite $N$ and to look for
estimation schemes which guarantee at least that error probabilities vanish ``as fast as possible'' as $N$
goes to infinity (cf. \cite{Helstrom,HolBook}; for a collection of recent publications on the subject see also
\cite{HayashiBook}). There are two approaches which implement this somewhat vague idea in a  mathematically
exact way. One possibility is to look at variances (rescaled by $N$) in the limit $N \to \infty$. This is done in
several works (cf. e.g. \cite{GillMassar00,Gill01,Matsumoto02} and in particular the papers reprinted in
\cite{HayashiBook}) and it leads to quantum analogs of classical Cramer-Rao type bounds. The second idea is to
analyze the large deviation behavior of the estimators. To make this more precise let us denote an estimate
derived from a measurement on $N$ systems in the joint state $\rho^{\otimes N}$ by $\sigma$. Then we can look at the
probability $P_{N,\epsilon}$ that the trace-norm distance between $\rho$ and $\sigma$ (or any other appropriate distance
measure for states) is at least $\epsilon$, i.e. $\|\rho - \sigma\|_1 \geq \epsilon$. Since $\rho=\sigma$ would be the exact estimate this is
clearly an error probability. Now we are interested in those cases where $P_{N,\epsilon}$ vanishes exponentially fast
in $N$, i.e.  
\begin{equation} \label{eq:1}
  P_{N,\epsilon}  \approx C_N \exp\bigl( -N \inf_{\|\rho - \sigma\|_1 \geq \epsilon} \, I(\sigma,\rho) \bigr).
\end{equation}
Here $C_N$, $N \in \Bbb{N}$ is an unknown sequence of positive real numbers, growing at most subexponentially with
$N$ (and which is of no interest for the following), and $I(\rho,\sigma)$ is a positive function which vanishes iff $\sigma
= \rho$ holds. $I$ is called the \emph{rate function} because it describes the exponential rate with which
estimation errors vanishes asymptotically. In classical statistics this analysis was initiated by Bahadur 
\cite{MR0293767,MR0207085,MR0315820} and has become in the mean time a classical topic (``Bahadur
efficiency''). About the quantum case, however, much less is known, and the results available so far cover
three different areas: 1. In \cite{KWEst,AlRuSa88,MR1955142} an explicit scheme to estimate the spectrum of $\rho$
is proposed and its rate function is calculated. The latter is shown to be optimal in \cite{MR1947128}. 2. The
rate function of the optimal pure state estimator is calculated in \cite{Hayashi98}. 3. In \cite{MR1947128}
the behavior of quantities like $\lim_{\epsilon\to0} \inf_{\|\rho - \sigma\|_1 \geq \epsilon} I(\rho,\sigma)$ is analyzed for one-parameter
families of states, and the relation to  quantum Fisher information is discussed.  

The purpose of the present paper is to extend the results about the spectrum in \cite{KWEst} and about pure
states in \cite{Hayashi98} in two respects. Firstly, we will propose a scheme to estimate the full density
matrix which is based on covariant observables \cite{HolBook} and which reduces to \cite{KWEst} if we look
only at the spectrum of $\rho$. And secondly, we will pose the question whether the proposed scheme is
``asymptotically optimal'', i.e. whether its rate function is bigger than the rate function of any other
scheme. There is of course no guarantee that a given set of functions admits a maximal element, but in the
classical case it is known that such an ``optimal rate function'' exists (and is given by the classical
relative entropy -- this is again a consequence of Bahadur's work \cite{MR0293767,MR0207085,MR0315820}). For
quantum systems, however, the situation is -- not very surprisingly -- much more difficult.  

The outline of the paper is as follows: In Section \ref{sec:stat-probl-main} we will give a more formal
introduction to the questions we are considering and in \ref{sec:stat-main-results} we will state our main
results. The proofs and a more detailed discussion is then distributed among Section \ref{sec:full-problem}
(were we will consider $\U(d)$-covariant estimation schemes) and Section \ref{sec:upper-bounds} (where upper
bounds on rate functions will be discussed).

%%%%%%%%%%%%%%%%%%%%%%%%%%%%%%%%%%%%%%%%%%%%%%%%%%%%%%%%%%%%%%%%%%%%%%%%%%%%%%%%%%%%%%%%%%%%%%%%%%%%
\section{Basic definitions}
\label{sec:stat-probl-main}
%%%%%%%%%%%%%%%%%%%%%%%%%%%%%%%%%%%%%%%%%%%%%%%%%%%%%%%%%%%%%%%%%%%%%%%%%%%%%%%%%%%%%%%%%%%%%%%%%%%%

In this section we will present some mathematical preliminaries (in particular basic definitions and
terminology) concerning quantum state estimation. A short summary of material from the theory of large
deviations used throughout this paper can be found in Appendix \ref{sec:some-material-from}.

%%%%%%%%%%%%%%%%%%%%%%%%%%%%%%%%%%%%%%%%%%%%%%%%%%
\subsection{State estimation}
\label{sec:state-estimation}
%%%%%%%%%%%%%%%%%%%%%%%%%%%%%%%%%%%%%%%%%%%%%%%%%%

Let us consider the $d$-dimensional Hilbert space $\scr{H} = \Bbb{C}^d$ and the corresponding set $\scr{S}$ of
density operators. The task of quantum state estimation is to determine a state $\rho \in \scr{S}$ by a measurement
on an $N$-fold system, which is prepared in the joint state $\rho^{\otimes   N}$. Mathematically this can be described
by a normalized POV measure $E_N$ on the state space $\scr{S}$ with values in the algebra $\scr{B}(\scr{H}^{\otimes
  N})$ of (bounded) operators on $\scr{H}^{\otimes N}$. More precisely, $E_N$ is a (strongly) \emph{$\sigma$-additive}
set function 
\begin{equation}
  E_N: \goth{B}(\scr{S}) \to \scr{B}(\scr{H}^{\otimes N})\ \text{with}\ E_N(\Delta) \geq 0,\ E_N(\emptyset)=0,\ E_N(\scr{S}) = \Bbb{1},
\end{equation}
on the \emph{Borel $\sigma$ algebra} $\goth{B}(\scr{S})$ of $\scr{S}$, and the probability to get an estimate in a 
Borel set $\Delta \subset \scr{S}$ is given by   
\begin{equation} \label{eq:3}
  \mu_{N,\rho}(\Delta) = \tr\bigl(\rho^{\otimes N} E_N(\Delta) \bigr).
\end{equation}
Since the number $N$ of systems is arbitrary, we need a whole sequence of observables and we will call each
such sequence in the following a \emph{full estimation scheme}. For a good estimation scheme the quality of 
the estimates should increase with $N$, i.e. the error probability should decrease and in the limit of
infinitely many input systems the estimate should be exact; in other words the sequence of probability measures
$(\mu_{N,\rho})_{N \in \Bbb{N}}$ should converge for each $\rho$ weakly to the point measure concentrated at $\rho$. Such
an estimation scheme is called \emph{consistent}.

If we are interested not in the whole state but only in some special properties of $\rho$ (e.g. its von Neumann
entropy), described by a function $\scr{S} \ni \rho \mapsto p(\rho) \in X$ taking its values in a locally compact, separable
metric space $X$ we have to consider more generally POV measures $E_N : \goth{B}(X) \to \scr{B}(\scr{H}^{\otimes N})$
on $X$ instead of $\scr{S}$. As before, $\tr\bigl( \rho^{\otimes N} E_N(\Delta) \bigr)$ is the probability to get an
estimate in $\Delta \subset X$. Estimating the spectrum of a density operator is a particular example of this kind. In
this case $p$ coincides with 
\begin{equation} \label{eq:21}
  s: \scr{S} \to \Sigma = \{ x \in [0,1]^d \, | \, x_1  \geq \cdots \geq x_d \geq 0, \sum_{j=1}^d x_j = 1\}
\end{equation}
which maps a density operator $\rho$ to its spectrum $s(\rho) \in \Sigma$, i.e. $s_j(\rho) = \langle\chi_j,\rho\chi_j\rangle$ where $\chi_1, \ldots, \chi_d$
denotes an appropriate eigenbasis of $\rho$. We will call $\Sigma$ the \emph{set of ordered spectra} and $s$ the 
\emph{canonical projection} onto $\Sigma$. Let us summarize the discussion up to now in the following definition.

\begin{defi} \label{def:2}
  Consider a finite dimensional Hilbert space $\scr{H}$, the corresponding set $\scr{S}$ of density operators,
  and a function $p : \scr{S} \to X$ taking its values in the locally compact, separable metric space $X$. A
  sequence $(E_N)_{N \in \Bbb{N}}$ of POV measures $E_N: \goth{B}(X) \to \scr{B}(\scr{H}^{\otimes N})$ is called a
  \emph{$p$-estimation scheme} (or just an estimation scheme if there is no danger of confusion). A
  $p$-estimation scheme is called \emph{consistent}, if the sequence $(\mu_{N,\rho})_{N \in \Bbb{N}}$ of probability
  measures defined in (\ref{eq:3}) 
%  \begin{equation} \label{eq:7}
%    \mu_{\rho,N}(\Delta) = \tr \bigl(\rho^{\otimes N} E_N(\Delta)\bigr)
%  \end{equation}
  converges for each $\rho \in \scr{S}$ weakly to a point measure concentrated at $p(\rho) \in X$.
\end{defi}

We recover both cases we are mainly interested in if we set $X = \scr{S}$ and $p=\Id$ for the full problem and
$X=\Sigma$ and $p=s$ for spectral estimation. 

Of special importance in this work are estimation scheme with additional symmetry properties: Let us denote
the permutation group on $N$ points by $\Sym_N$ and its natural representation on $\scr{H}^{\otimes N}$ by $V$, i.e.
\begin{equation} \label{eq:112}
  V_\sigma \psi_1 \otimes \cdots \otimes \psi_N = \psi_{\sigma^{-1}(1)} \otimes \cdots \otimes \psi_{\sigma^{-1}(N)},\quad \sigma \in \Sym_N,\ \psi_1, \ldots, \psi_N \in \scr{H}.
\end{equation}
An estimation scheme $(E_N)_{N \in \Bbb{N}}$ is called \emph{permutation invariant}, if
\begin{equation} \label{eq:113}
  V_\sigma E_N(\Delta) V_\sigma^* = E_N(\Delta)\quad \forall \sigma \in \Sym_N\ \forall \Delta \in \goth{B}(X)
\end{equation}
holds. Likewise, it is called \emph{$\U(d)$-covariant} (or just covariant) if $\U(d)$ acts continuously on $X$
by $\U(d) \times X \ni (U,x) \mapsto \alpha_U(x) \in X$ such that the conditions
\begin{equation} \label{eq:110}
  U^{\otimes N} E_N(\Delta) U^{\otimes N *} = E_N\bigl(\alpha_U(\Delta)\bigr)\quad \forall U \in \U(d)\ \forall \Delta \in \goth{B}(X)
\end{equation}
and
\begin{equation} \label{eq:111}
  p(U\rho U^*) = \alpha_U\bigl(p(\rho)\bigr)\quad \forall U \in \U(d)\ \forall \rho \in \scr{S}
\end{equation}
are satisfied. If the scheme $(E_N)_{N \in \Bbb{N}}$ is consistent, covariance of the projection $p$
(\ref{eq:111}) is implied by covariance of the measures $E_N$ (\ref{eq:110}). Furthermore, note that the
$\U(d)$ operation $\alpha_U$ is uniquely determined (if it exists) due to surjectivity of $p$. For full estimation
we have $\alpha_U(\rho) = U\rho U^*$ and for spectral estimation it is the trivial action, i.e. $\alpha_U(x) = x$. Hence,
covariant estimation schemes are defined in both cases we are interested in.

%%%%%%%%%%%%%%%%%%%%%%%%%%%%%%%%%%%%%%%%%%%%%%%%%%
\subsection{Large deviations}
\label{sec:large-deviations}
%%%%%%%%%%%%%%%%%%%%%%%%%%%%%%%%%%%%%%%%%%%%%%%%%%

Consider now, a Borel set $\Delta \subset X$ and a state $\rho \in \scr{S}$ such that $p(\rho) \not\in \bar \Delta$ (the closure of
$\Delta$). The quantity $\mu_{\rho,N}(\Delta)$ is then the probability to get a false estimate in $\Delta$. If the scheme is
consistent this probability goes to zero. This is, however, a very weak statement because the convergence can
be very slow. As already pointed out in the introduction, we are therefore interested in 
schemes, where convergence of error probabilities to zero is \emph{exponentially fast}; in other words for
each $\rho \in \scr{S}$ the sequence $(\mu_{N,\rho})_{N \in \Bbb{N}}$ of probability measures from Equation (\ref{eq:3})
should satisfy the \emph{large deviation principle}\footnote{A short summary of definitions and theorems from
  large deviations theory which are relevant for this paper can be found in Appendix
  \ref{sec:some-material-from}.} with a rate function $I(\rho,\,\cdot\,)$. This idea leads to the following
definition:  

\begin{defi} \label{def:4}
  A $p$-estimation scheme $(E_N)_{N \in \Bbb{N}}$, as described in Definition \ref{def:2}, satisfies the
  \emph{large deviation principle} (LDP) with rate function $I: \scr{S} \times X \to [0,\infty]$ if     
  \begin{enumerate}
  \item 
    $I_\rho = I(\rho,\,\cdot\,)$ is a rate function (cf. Definition \ref{def:1}) for each $\rho \in \scr{S}$.
  \item \label{item:1}
    $I(\rho,x) = 0$ iff $p(\rho) = x$ holds.
  \item 
    The sequence $(\mu_{N,\rho})_{N \in \Bbb{N}}$ of probability measures (\ref{eq:3}) satisfies for each $\rho \in
    \scr{S}$ the large deviation principle with rate function $I_\rho$.  
  \end{enumerate}
\end{defi}

Note that condition \ref{item:1} guarantees that each scheme which satisfies the LDP is consistent, because 
the $\mu_{N,\rho}(\Delta)$ converge to $0$, if $\Delta$ is a closed set which does no contain $p(\rho)$. Occasionally we will
have to refer to the rate function $I$ of an estimation scheme $(E_N)_{N \in \Bbb{N}}$ without using $(E_N)_{N \in
  \Bbb{N}}$ directly. In this case we will call $I$ an admissible rate function.

\begin{defi}
  A function $I: \scr{S} \times X \to [0,\infty]$ which is the rate function of a $p$-estimation scheme is called
  \emph{$p$-admissible} (or just admissible if $p$ is understood).  The set of all $p$-admissible rate
  functions is denoted by $\scr{E}(p)$. 
\end{defi}

We do not yet know how continuous or discontinuous admissible rate functions can be in their first
argument. E.g. an otherwise very bad estimation scheme might provide very fast exponential decay for a
particular input state. The discussion in Sections \ref{sec:cont-prop} and \ref{sec:beyond-covariance} will
indicate that discontinuities might occur in particular at the boundary of the state space, while the
behavior in the interior of $\scr{S}$ (i.e. at non-degenerate density matrices) seems to be more regular. To
avoid such difficulties let us introduce the following subset of $\scr{E}(p)$:
\begin{equation} \label{eq:59}
  \scr{E}^0(p) = \{ I \in \scr{E}(p)\, | \, I\ \text{is lower semi-continuous}\, \}.  
\end{equation}
If the map $p$ we want to estimate is covariant in the sense of Equation (\ref{eq:111}) we can introduce in
addition 
\begin{equation} \label{eq:39}
  \scr{E}^c(p) = \{ I \in \scr{E}(p) \, | \, \text{$I$ is covariant}\, \}, 
\end{equation}
where we call an \emph{admissible rate function} \emph{covariant}, if it is the rate function of a
$\U(d)$-covariant estimation scheme. In contrast to this, \emph{any function} $F: \scr{S} \times X \to [0,\infty]$ is
called \emph{$\U(d)$-invariant} if 
\begin{equation} \label{eq:48}
  F\bigl(U\rho U^*,\alpha_U(x)\bigr) = F(\rho,x)\ \forall U \in \U(d)\ \forall \rho \in \scr{S}\ \forall x \in X 
\end{equation}
is satisfied. Obviously, each admissible rate function which is covariant is $\U(d)$-invariant too. It is not
clear whether the converse holds as well (i.e. whether $\U(d)$-invariance of $I \in \scr{E}(p)$ implies
covariance). However, problems can occur only on the boundary of $\scr{S}$ (i.e. for degenerate density
matrices) and even there only if $I$ is not lower semicontinuous (cf. Section \ref{sec:averaging} for 
details).  Finally note that $\U(d)$-invariance of $I \in \scr{E}^c(p)$ implies, together with lower
semi-continuity of $I_\rho(\,\cdot\,) = I(\rho,\,\cdot\,)$, lower semi-continuity of $I^x(\,\cdot\,) = I(\,\cdot\,,x)$ \emph{along
  the orbits} of the $\U(d)$ action on $\scr{S}$. The general relation between $\scr{E}^0(p)$ and
$\scr{E}^c(p)$ is, however, not clear (i.e. $I \in \scr{E}^c(p)$ can be discontinuous transversal to the
orbits).  

Ideally, we would like to have estimation schemes $(E_N)_{N \in \Bbb{N}}$ which provide the \emph{fastest
  possible exponential decay} of error probabilities. Hence, for a given map $p: \scr{S} \to X$ we are mainly
interested in the quantities   
\begin{equation} \label{eq:22}
  \scr{I}_p(\rho,\sigma) = \sup_{I \in \scr{E}(p)} I(\rho,\sigma),\quad \scr{I}_p^0(\rho,\sigma) = \sup_{I \in \scr{E}^0(p)} I(\rho,\sigma)\
\end{equation}
and
\begin{equation} \label{eq:34}
  \scr{I}_p^c(\rho,\sigma) = \sup_{I \in \scr{E}^c(p)} I(\rho,\sigma). 
\end{equation}
The functions $\scr{I}_p^\#: \scr{S} \times X \to [0,\infty]$ thus defined (following the notation introduced above, we
will write $\scr{I}_{\Id}^\#$ for full and $\scr{I}_s^\#$ for spectral estimation), are the least upper bounds
on the sets $\scr{E}^\#(p)$, but they are not necessarily admissible themselves. In slight abuse of language we
will call them nevertheless the \emph{optimal rate functions}. If $\scr{I}_p$ can be realized as the rate
function of a particular estimation scheme $(E_N)_{N \in \Bbb{N}}$, we will call $(E_N)_{N \in \Bbb{N}}$
(strongly) \emph{asymptotically optimal}.   

%%%%%%%%%%%%%%%%%%%%%%%%%%%%%%%%%%%%%%%%%%%%%%%%%%%%%%%%%%%%%%%%%%%%%%%%%%%%%%%%%%%%%%%%%%%%%%%%%%%%
\section{Summary of main results}
\label{sec:stat-main-results}
%%%%%%%%%%%%%%%%%%%%%%%%%%%%%%%%%%%%%%%%%%%%%%%%%%%%%%%%%%%%%%%%%%%%%%%%%%%%%%%%%%%%%%%%%%%%%%%%%%%%

A particular example for asymptotic optimality arises in classical estimation theory (for finite
probability distributions). It is known from Bahadur efficiency \cite{MR0293767,MR0207085,MR0315820} that the
classical relative entropy is  an upper bound for all admissible rate functions; and Sanov's theorem
(cf. eg. \cite{MR1739680}) states that this bound can be achieved by the empirical distribution (i.e. relative
frequencies in a given  sample). The latter provides therefore an asymptotically optimal estimation
scheme. For quantum systems the situation is more difficult, and our knowledge is
(unfortunately) not yet as complete as for classical estimation. Nevertheless, we have some significant
partial results which we want to summarize in this section. The proofs and a more detailed discussion are
postponed to Section \ref{sec:full-problem} and \ref{sec:upper-bounds}.

%%%%%%%%%%%%%%%%%%%%%%%%%%%%%%%%%%%%%%%%%%%%%%%%%%
\subsection{Estimating the spectrum}
\label{sec:estimating-spectrum-1}
%%%%%%%%%%%%%%%%%%%%%%%%%%%%%%%%%%%%%%%%%%%%%%%%%%

The most complete result is available for spectral estimation. To state it let us recall the definition of the
scheme presented in \cite{KWEst}. It is based on the decomposition of the representation $U \mapsto U^{\otimes N}$ of the
unitary group $\U(d)$ into irreducible components. The latter is given by  
\begin{equation}
  \scr{H}^{\otimes N} = \bigoplus_{Y \in \scr{Y}_d(N)} \scr{H}_Y \otimes \scr{K}_Y,\quad U^{\otimes N} = \bigoplus_{Y \in \scr{Y}_d(N)} \pi_Y(U) \otimes \Bbb{1},
\end{equation}
where $\scr{Y}_d(N)$ denotes the set of \emph{Young frames} with $d$ rows and $N$ boxes
\begin{equation}
  \scr{Y}_d(N) = \{ Y \in \Bbb{N}^d\,|\, Y_1 \geq \cdots \geq Y_d,\ \sum_{j=1}^d Y_j = N \}, 
\end{equation}
$\pi_Y$ denotes the irreducible representation with \emph{highest weight}\footnote{More precisely the $Y_1, \ldots,
  Y_d$ are the components of the highest weight in a particular basis of the Cartan subalgebra.} $Y$, and
$\scr{K}_Y$ is a multiplicity space which carries an irreducible representation of the symmetric group $\Sym_N$
on $N$ elements: 
\begin{equation}
  V_\sigma = \bigoplus_{Y \in \scr{Y}_d(N)} \Bbb{1} \otimes \Pi_Y(\sigma),\quad \sigma \in \Sym_N 
\end{equation}
where $V_\sigma$ is defined in Equation  (\ref{eq:112}) and $\Pi_Y$ is the irreducible $\Sym_N$ representation
defined by the Young frame $Y$.  

Now we can define a \emph{spectral estimation scheme} $(\hat{F}_N)_{N \in \Bbb{N}}$ by 
\begin{equation} \label{eq:4}
  \hat{F}_N (\Delta) = \sum_{Y/N \in \Delta} P_Y, 
\end{equation}
where $P_Y$ denotes the projection onto $\scr{H}_Y \otimes \scr{K}_Y$:
\begin{equation} \label{eq:58}
  P_Y \in \scr{B}(\scr{H}^{\otimes N}),\ P_Y^2 = P_Y,\ P_Y^* = P_Y,\ P_Y \scr{H}^{\otimes N} = \scr{H}_Y \otimes \scr{K}_Y.
\end{equation}
In other words $\hat{F}_N$ is a discrete measure with \emph{normalized} Young frames $Y/N$ as possible
estimates and the probability to get the outcome $Y/N$ for input systems in the joint state $\rho^{\otimes N}$ is
$\tr(\rho^{\otimes N} P_Y)$. In \cite{KWEst} it is shown that $\hat{F}_N$ satisfies the large deviation principle with the
classical relative entropy between the probability vectors $x \in \Sigma$ and $s(\rho)$ as the rate function $I(\rho,x)$. As we
will see in Subsection \ref{sec:proof-theor-refthm:1} this is in fact the best that can be achieved (cf. also
\cite{MR1947128}).

\begin{thm} \label{thm:5}
  The spectral estimation scheme $(\hat{F}_N)_{N \in \Bbb{N}}$ defined in (\ref{eq:4}) is 
  asymptotically optimal; i.e. it satisfies the LDP with the optimal rate function $\scr{I}_s$ defined in
  Equation (\ref{eq:22}). In addition $\scr{I}_s = \scr{I}_s^0 = \scr{I}_s^c$ holds, and $\scr{I}_s$ is
  given explicitly by  
  \begin{equation}
    \scr{S} \times \Sigma \ni (\rho,x) \mapsto \scr{I}_s(\rho,x) = \sum_{j=1}^d x_j\bigl[\ln(x_j) - \ln\bigl(s_j(\rho)\bigr)\bigr].
  \end{equation}
  where $s: \scr{S} \to \Sigma$ is the canonical projection from Equation (\ref{eq:21}).
\end{thm}

%%%%%%%%%%%%%%%%%%%%%%%%%%%%%%%%%%%%%%%%%%%%%%%%%%
\subsection{The full density matrix}
\label{sec:full-density-matrix}
%%%%%%%%%%%%%%%%%%%%%%%%%%%%%%%%%%%%%%%%%%%%%%%%%%

For the full problem the best scheme $(\hat{E}_N)_{N \in \Bbb{N}}$ we have found so far is defined by the
integral (with an arbitrary continuous function $f: \scr{S} \to \Bbb{R}$)  
\begin{multline} \label{eq:8} 
  \int_\scr{S} f(\rho) \hat{E}_N(d\rho) = \\ \sum_{Y \in \scr{Y}_d(N)} \dim \scr{H}_Y \int_{\U(d)} f(U \rho_{Y/N} U^*)\,\,
   \kb{\pi_Y(U) \phi_Y} \otimes \Bbb{1}\,\, dU,
\end{multline}
where $\phi_Y \in \scr{H}_Y$ is the \emph{highest weight vector} of the irreducible representation $\pi_Y$ and
$\rho_x$ denotes for each $x \in \Sigma$  the diagonal density matrix
\begin{equation} \label{eq:9}
  \rho_x = \diag(x_1,\ldots,x_d).
\end{equation}
The main properties of this scheme are: It \emph{projects} to the spectral estimation scheme $\hat{F}_N$ from 
Subsection \ref{sec:estimating-spectrum-1}
\begin{equation} \label{eq:5}
  \hat{E}_N\bigl(s^{-1}(\Delta)\bigr) = \hat{F}_N(\Delta)\quad \forall \Delta \in \goth{B}(\Sigma),
\end{equation}
it is \emph{covariant} (i.e. Equation (\ref{eq:110}) holds with $\alpha_U(\rho) = U\rho U^*$) and \emph{permutation
  invariant} (cf. Equation (\ref{eq:113})).
Measuring $\hat{E}_N$ can be regarded therefore as a two step process: First measure the observable
$\hat{F}_N$ in terms of the instrument $T$, which is defined by the family of channels
(given in the Schr\"odinger picture):
\begin{equation}
  T_Y : \scr{B}(\scr{H}^{\otimes N}) \ni \omega \mapsto \tr_{K_Y} ( P_Y \omega P_Y) \in \scr{B}(\scr{H}_Y) \quad Y \in \scr{Y}_d(N),
\end{equation}
where $\tr_{K_Y}$ denotes the partial trace over $\scr{K}_Y$ and the $P_Y$ are again the projections from
(\ref{eq:58}). If the estimate for the spectrum we get in this way (with probability $\tr(P_Y\rho^{\otimes N})$) is
$Y/N$, the output of $T$ is a quantum system (described by the Hilbert space $\scr{H}_Y$ -- hence of different
type then the input system\footnote{If $d=2$ holds the situation is special. In this case the output of $T$
  can be regarded as an $M = Y_1 - Y_2$ qubit system, and $T$ itself coincides with the ``natural purifier''
  studied in \cite{CEM99,pur}.}) in the state $\tr(P_Y \rho^{\otimes N})^{-1} T_Y(\rho^{\otimes N})$. On this system we perform a
measurement of a covariant observable $E_Y$ with values in $\scr{S}_Y = s^{-1}(Y/N)$ which is defined by the
integral  
\begin{equation} \label{eq:23}
  \int_{\scr{S}_Y} f(\sigma) E_Y(d\sigma) = \int_{\U(d)}  f(U \rho_{Y/N} U^*) \,\, \kb{\pi_Y(U)\phi_Y} \,\, dU,
\end{equation}
(where $f$ denotes now a continuous function on $\scr{S}_Y$) and this gives us an estimate for the
\emph{eigenvectors} of $\rho$. In the special case of pure states (i.e. if the first measurement gives $Y/N = (1,
0, 0, \ldots, 0)$) the observable $\hat{E}_Y$ is given by 
\begin{equation}
  \int_\scr{P} f(\sigma) \hat{E}_Y(d\sigma) = \int_\scr{P} f(\sigma) \sigma^{\otimes N},\quad \text{for}\ Y = (N,0,\ldots,0),
\end{equation}
where $\scr{P} = s^{-1}(1,0,\ldots,0)$ denotes the set of pure states. This observable is known to optimize for
each $N$ global quality criteria like averaged fidelity \cite{HolBook,MasPop95,Hayashi98}. Hence we can look at
$\hat{E}_N$ as a direct generalization of the best known  estimation schemes for the spectrum and for pure
states. We discuss this point of view in greater detail in Section \ref{sec:an-explicit-scheme}. The large
deviation behavior of $\hat{E}_N$ is described by the following theorem (cf. Section
\ref{sec:proof-theor-refthm:3} for a proof):   

\begin{thm} \label{thm:3}
  The full estimation scheme $(\hat{E}_N)_{N \in \Bbb{N}}$ defined in Equation (\ref{eq:8}) satisfies the large
  deviation principle with rate function $\hat{I}:\scr{S} \times \scr{S} \to [0,\infty]$ 
  \begin{equation} \label{eq:11}
     \hat{I}(\rho,U\rho_xU^*) = \sum_{k=1}^d \left(x_k \ln(x_k) - (x_k - x_{k+1}) \ln\bigl[\PM_k(U^*\rho U)\bigr]\right) 
  \end{equation}
  where $x = (x_1,\ldots,x_d) \in \Sigma$, $x_{d+1}=0$, $\rho_x$ is the density matrix from Equation (\ref{eq:9}), $U \in
  \U(d)$, and $\PM_j(\sigma)$ denotes the \emph{principal minor} (i.e. the upper left rank $j$ subdeterminant) of
  the matrix $\sigma$. 
\end{thm}

The best upper bound on the rate function for full estimation schemes we have found so far is derived from
quantum hypothesis testing. %(cf. Subsection \ref{sec:proof-theor-refthm:1} and the references therein). 

\begin{thm} \label{thm:1}
  Each admissible rate function $I: \scr{S} \times \scr{S} \to [0,\infty]$ is bounded from above by the relative entropy,
  i.e. 
  \begin{equation}
    I(\rho,\sigma) \leq S(\rho,\sigma) = \tr\bigl(\sigma\ln(\sigma) - \sigma\ln(\rho)\bigr)\quad \forall \rho,\sigma \in \scr{S}.
  \end{equation}
\end{thm}

\begin{figure}[htbp]
  \begin{center}

    \includegraphics{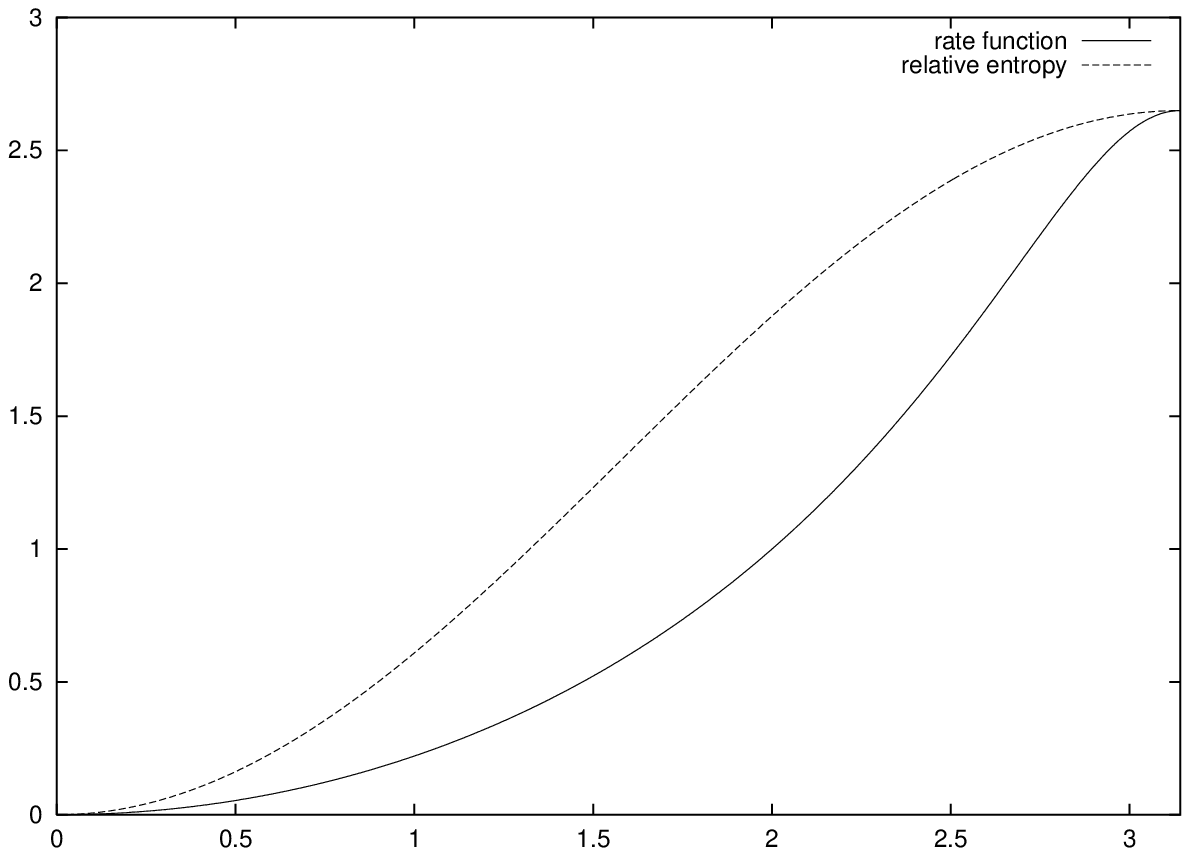}

\includegraphics[scale=1.0]{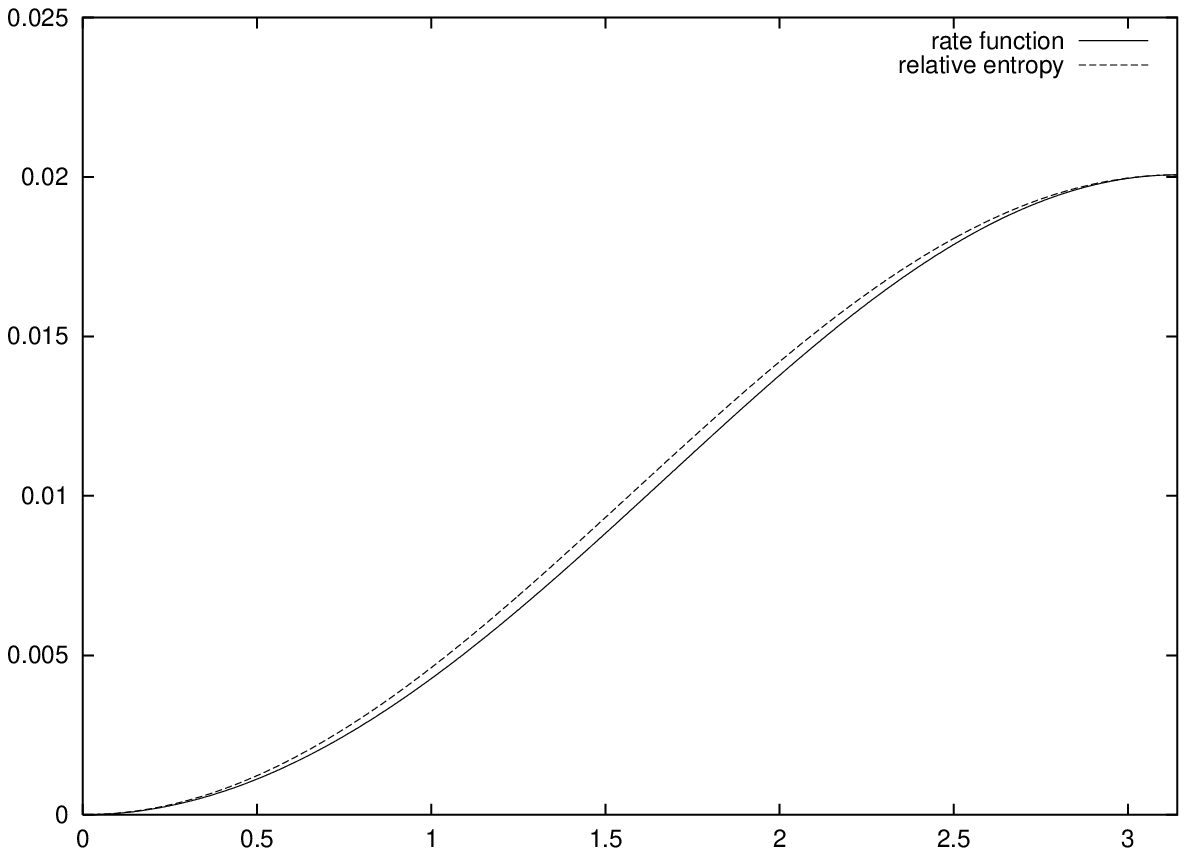}
  \end{center}
  
  \caption{\label{fig:1}
    Relative entropy and rate function $\hat{I}$ as a function of the angle $\theta$ between the two Bloch
    vectors $\vec{x}$ and $\vec{y}$. The upper plot corresponds to the case $|\vec{x}| = |\vec{y}| = 0.9$ and
    the lower to $|\vec{x}| = |\vec{y}| = 0.1$.}
\end{figure}

The proof will be given in Section \ref{sec:proof-theor-refthm:1}; cf also \cite{MR1947128}. It is easy to
check numerically that $\hat{I}(\rho,\sigma)$ and $S(\rho,\sigma)$ do not coincide in general. If we consider in particular
the qubit case ($d=2$) and express the density operators $\rho,\sigma$ in Bloch form, i.e.  
\begin{equation}
  \rho = \frac{1}{2} \bigl[ \Bbb{1} + \vec{x} \cdot \vec{\sigma}\bigr],\ \sigma = \frac{1}{2} \bigl[ \Bbb{1} + \vec{y} \cdot
  \vec{\sigma}\bigr]  
\end{equation}
(where $\vec{\sigma} = (\sigma_1,\sigma_2,\sigma_3)$ are the Pauli matrices and $\vec{x},\vec{y} \in \Bbb{R}^3$ with $|\vec{x}|,
|\vec{y}| \leq 1$), we get for the rate function $I$ from Equation (\ref{eq:11}) 
\begin{equation}
  \hat{I}(\rho,\sigma) = -S(\sigma) - |\vec{y}| \ln\left[ \frac{1+|\vec{x}| \cos\theta}{2}\right] -
  \frac{1-|\vec{y}|}{2}\ln\left[\frac{1-|\vec{x}|^2}{4}\right],
\end{equation}
where $\theta$ denotes the angle between $\vec{x}$ and $\vec{y}$ and $S(\sigma)$ is the von Neumann entropy of $\sigma$. The
relative entropy of $\sigma$ and $\rho$ becomes \cite{Cortese02} 
\begin{equation}
  S(\rho,\sigma) = -S(\sigma) - \frac{1}{2} \ln(1+|\vec{x}|^2) - \frac{|\vec{y}| \cos(\theta)}{2} \ln\left(
    \frac{1+|\vec{x}|}{1-|\vec{x}|} \right).
\end{equation}
We have plotted both quantities as functions of $\theta$ for two different values of $|\vec{x}| = |\vec{y}|$ in
Figure \ref{fig:1}, which shows that $I(\rho,\sigma)$ is in general strictly smaller than $S(\rho,\sigma)$.

%%%%%%%%%%%%%%%%%%%%%%%%%%%%%%%%%%%%%%%%%%%%%%%%%%%
\subsection{Optimal rate functions}
\label{sec:optim-rate-funct}
%%%%%%%%%%%%%%%%%%%%%%%%%%%%%%%%%%%%%%%%%%%%%%%%%%%

Hence, for a general input state $\rho$ we only know for sure that the optimal rate functions defined in
Equation (\ref{eq:22}) and (\ref{eq:34}) have to satisfy (with $p=\Id$ for full estimation)
\begin{equation} \label{eq:27}
  \hat{I} \leq \scr{I}_{\Id}^c,\scr{I}_{\Id}^0 \leq \scr{I}_{\Id} \leq S.
\end{equation}
This is, however, not as bad as it looks like at a first glance: Since $S(\rho,\sigma)$ and $\hat{I}(\rho,\sigma)$ coincide if
$\rho$ and $\sigma$ commute, we get 
\begin{multline}
  \hat{I}(\rho,\sigma) = \scr{I}_{\Id}^c(\rho,\sigma) = \scr{I}_{\Id}^0(\rho,\sigma) = \scr{I}_{\Id}(\rho,\sigma) = S(\rho,\sigma) = \\ 
  \sum_{j=1}^d s_j(\sigma)\, \bigl( \ln s_j(\sigma) - \ln s_j(\rho) \bigr)\quad \forall \rho,\sigma \in \scr{S}\ \text{with}\ [\rho,\sigma] = 0.
\end{multline}
A second partial result arises if the input state is pure. In Proposition \ref{thm:10} we will show
\begin{equation} \label{eq:36}
  \scr{I}^c_{\Id}(\rho,\sigma) = \hat{I}(\rho,\sigma)\quad \forall \rho,\sigma \in \scr{S}\ \text{with $\rho$ pure},
\end{equation}
and in Section \ref{sec:an-explicit-scheme} we will give some heuristic arguments which indicate that
$\hat{I}$ and $\scr{I}^c$ coincide even for general input states. This indicates that $(\hat{E}_N)_{N \in
  \Bbb{N}}$ is the best scheme as long as we are insisting on some additional regularity conditions of the
rate function -- in the case at hand this is covariance. It is not clear, however, whether covariance can be
replaced by something more general without breaking the equality with $\hat{I}$. There are at least some
indications (cf. Section \ref{sec:pure-states}) that Equation (\ref{eq:36}) would still hold if we replace 
$\scr{I}^c_{\Id}$ with $\scr{I}^0_{\Id}$. Note that $\hat{I} \in \scr{E}^0(p)$ hence (\ref{eq:36}) already
implies $\scr{I}^0(\rho,\sigma) \geq \scr{I}^c(\rho,\sigma)$ for pure $\rho$. Our conjecture here is that equality holds for all $\rho$
and $\sigma$.   

Another result which can be derived easily  from Equation (\ref{eq:36}) and Proposition \ref{prop:1} is
$\scr{S} \not\in \scr{E}(\Id)$, i.e. there is no estimation scheme with relative entropy as its rate
function. This follows from the fact that $S$ is lower semicontinuous and $\U(d)$-invariant in the sense of
Equation (\ref{eq:48}). Hence $S \in \scr{E}(\Id)$ would imply according to Proposition \ref{prop:1} $S \in
\scr{E}^c(\Id)$ in contradiction to Equation (\ref{eq:36}) and the fact that $S(\rho,\sigma) > \hat{I}(\rho,\sigma)$ holds for
all pure states $\rho,\sigma$ with $\rho \neq \sigma$ and $\rho\sigma \neq 0$. On the other hand there is strong evidence that
$\scr{I}_{\Id} = S$ holds, i.e. that $S$ is the best upper bound of the set of all admissible rate
functions. This would imply that we can find for each pair $\rho_0, \sigma_0 \in \scr{S}$ an $I \in \scr{E}(\Id)$ such
that $I(\rho_0,\sigma_0) = S(\rho_0,\sigma_0)$ holds, but $I$ is much smaller than $S$ (most probably even smaller than
$\hat{I}$) almost everywhere else. In Section \ref{sec:pure-states} we will discuss these topics in greater
detail. For now, let us summarize all our \emph{conjectures} in the following Equation 
\begin{equation} \label{eq:71}
  \hat{I} = \scr{I}^c_{\Id} = \scr{I}^0_{\Id} \leq \scr{I} = S. 
\end{equation}

%%%%%%%%%%%%%%%%%%%%%%%%%%%%%%%%%%%%%%%%%%%%%%%%%%%%%%%%%%%%%%%%%%%%%%%%%%%%%%%%%%%%%%%%%%%%%%%%%%%%
\section{Covariant observables}
\label{sec:full-problem}
%%%%%%%%%%%%%%%%%%%%%%%%%%%%%%%%%%%%%%%%%%%%%%%%%%%%%%%%%%%%%%%%%%%%%%%%%%%%%%%%%%%%%%%%%%%%%%%%%%%%

The aim of this section is to study estimation schemes which are $\U(d)$ covariant and permutation invariant, 
i.e. they do not prefer a special copy of the input state or a particular direction in the Hilbert space
$\scr{H}$. Among a proof of Theorem \ref{thm:3} we will provide several general results, which are useful
within the discussion of the questions raised in Section \ref{sec:optim-rate-funct}. Therefore only full
estimation schemes are considered in this section (i.e. $p=\Id$), but most of the results in Subsection 
\ref{sec:averaging} and \ref{sec:general-structure} can be generalized quite easily to $p$-estimation schemes,
if $p$ is sufficiently covariant.

%%%%%%%%%%%%%%%%%%%%%%%%%%%%%%%%%%%%%%%%%%%%%%%%%%
\subsection{Continuity properties}
\label{sec:cont-prop}
%%%%%%%%%%%%%%%%%%%%%%%%%%%%%%%%%%%%%%%%%%%%%%%%%%
 
Let us start with some technical results concerning continuity and uniform convergence with respect to the
original density matrix $\rho$. They will become crucial within the discussion of group averages in the next
section. Some of them, however, are quite interesting in their own right, and it is therefore reasonable to
devote a whole subsection for them.

Central subjects of this discussion will be integrals of the form 
\begin{equation} \label{eq:84}
  h_N(\rho,f) = \frac{-1}{N} \ln \int_\scr{S} e^{-N f(\sigma)} \tr\bigl( \rho^{\otimes N} E_N(d\sigma) \bigr),
\end{equation}
where $f$ denotes an arbitrary, real valued function on $\scr{S}$. Quantities of this form usually appear in
Varadhan's Theorem (cf. Theorem \ref{thm:7}), i.e. if the estimation scheme $(E_N)_{N \in \Bbb{N}}$ satisfies
the LDP with rate function $I$ we have 
\begin{equation} \label{eq:97}
  \lim_{N \to \infty} h_N(\rho,f) = h(\rho,f) = \inf_{\sigma \in \scr{S}} \bigl( I(\rho,\sigma) + f(\sigma) \bigr).
\end{equation}
If on the other hand $(E_N)_{N \in \Bbb{N}}$ does not necessarily satisfy the LDP but (\ref{eq:97}) holds for
each $f$ and a density matrix $\rho$, the sequence of probability measures $\tr\bigl(\rho^{\otimes N} E_N(\,\cdot\,)\bigr)$
satisfies the \emph{Laplace principle} (Definition \ref{def:5}) which is equivalent to the large deviation
principle (Theorem \ref{thm:8}). Hence the study of convergence properties of the $h_N(\rho,f)$ is a useful tool
to prove that the LDP holds for a given estimation scheme. 

In this section we will discuss continuity of $h$ with respect to $\rho$ and uniformity of the convergence $h_N \to
h$ (again with respect to $\rho$). The most crucial step in this direction is the following lemma.

\begin{lem} \label{lem:4}
  Consider an estimation scheme $(E_N)_{N \in \Bbb{N}}$ satisfying the LDP with rate function $I$, an
  arbitrary continuous (real valued) function $f$ and the functionals $h_N, h$ defined in Equations
  (\ref{eq:84}) and (\ref{eq:97}).
  \begin{enumerate}
  \item \label{item:2}
    For each non-degenerate density matrix $\rho \in \scr{S}$ and each sequence $\Bbb{N} \ni N \mapsto \rho_N \in \scr{S}$
    converging to $\rho$ we have
    \begin{equation} \label{eq:60}
      \lim_{N \to \infty} h_N(\rho_N,f) = \lim_{N \to \infty} h_N(\rho,f) = h(\rho,f) %= \inf_{\sigma \in \scr{S}} \bigl(I(\rho,\sigma) + f(\rho)\bigr)
    \end{equation}
  \item \label{item:3}
    If $I$ is lower semicontinuous \emph{in both arguments}, the lower bound
    \begin{equation} \label{eq:89}
       \liminf_{N \to \infty} h_N(\rho_N,f) \geq h(\rho,f)
    \end{equation}
    holds even for degenerate $\rho$.
  \end{enumerate}
\end{lem}

\begin{pf}
  Let us consider part \ref{item:2} first. In this case the proof mainly depends on the following lemma which
  allows us to represent one sequence as a convex combination of two others.

  \begin{lem} \label{lem:3}
    Consider two sequences $\Bbb{N} \ni N \mapsto \rho_N^{(j)} \in \scr{S}$, $j=1,2$ both converging to the same
    non-degenerate density matrix $\rho \in \scr{S}$. For each $\lambda \in \Bbb{R}$ with $0 < \lambda < 1$ there exists an
    integer $N_\lambda \in \Bbb{N}$ and a third sequence $\Bbb{N} \ni N \mapsto \sigma_N \in \scr{S}$ such that
    \begin{equation}
      \rho_N^{(1)} = \lambda \rho_N^{(2)} + (1-\lambda)\sigma_N\quad \forall N > \Bbb{N}_\lambda
    \end{equation}
    holds.
  \end{lem}

  \begin{pf}
    Let $\kappa  = \inf_{\|\phi\|=1} \langle\phi,\rho\phi\rangle$ and define 
    \begin{equation} \label{eq:61}
      \epsilon = \frac{(1-\lambda) \kappa}{\lambda+1}.
    \end{equation}
    Since $\rho$ is non-degenerate, we have $\kappa > 0$ and therefore $\epsilon>0$ as well. Hence there is an $N_\lambda \in
    \Bbb{N}$ such that (with $\phi \in \scr{H}$ and $A \in \scr{B}(\scr{H})$)
    \begin{align}
      \sup_{\|\phi\|=1}  | \langle\phi,(\rho_N^{(j)} - \rho) \phi\rangle | & \leq \sup_{\|A\|=1} \bigl|\tr\bigl( (\rho_N^{(j)} - \rho) A\bigr)\bigr| \\
      &= \|\rho_N^{(j)} - \rho\|_1 < \epsilon \label{eq:80}
    \end{align}
    holds for all $N > N_\lambda$ and for $j=1,2$. In addition we see by the triangle inequality that 
    \begin{equation} \label{eq:79}
      \sup_{\|\phi\|=1} | \langle\phi, (\rho_N^{(1)} - \rho_N^{(2)}) \phi\rangle | < 2 \epsilon
    \end{equation}
    holds as well for all $N > N_\lambda$. Now define 
    \begin{equation} \label{eq:70}
      \delta = \frac{\kappa}{2 \epsilon} - \frac{1}{2} = \frac{\lambda}{1-\lambda}
    \end{equation}
    (the second equality follows from Equation (\ref{eq:61})) and 
    \begin{equation}
      \sigma_N = - \delta \rho_N^{(2)} + (1+\delta) \rho_N^{(1)}\ \text{for}\ N > N_\lambda
    \end{equation}
    (and $\sigma_N \in \scr{S}$ arbitrary otherwise). Obviously $\tr(\sigma_N) = 1$ and
    \begin{equation}
      \frac{-\lambda}{1-\lambda} \rho_N^{(2)} + \frac{1}{1-\lambda} \rho_N^{(1)} = \sigma_N.
    \end{equation}
    Hence
    \begin{equation}
      \rho_N^{(1)} = \lambda \rho_N^{(2)} + (1-\lambda) \sigma_N\quad \forall N > N_\lambda
    \end{equation}
    as stated.

    It only remains to show that $\sigma_N \geq 0$ (and therefore $\sigma_N \in \scr{S}$) holds for all $N > N_\lambda$. This
    follows from
    \begin{align}
      \langle\phi, \sigma_N \phi\rangle &= - \delta \langle\phi,\rho_N^{(2)}\phi\rangle + (1+\delta) \langle\phi, \rho_N^{(1)}\phi\rangle \\
      & \geq -2 \delta \epsilon - \delta \langle\phi, \rho_N^{(1)} \phi\rangle + (1+\delta) \langle\phi, \rho_N^{(1)}\phi\rangle \label{eq:81}  \\
      &= -2 \delta \epsilon + \langle\phi, \rho_N^{(1)} \phi\rangle \geq -2 \delta \epsilon + \langle\phi, \rho \phi\rangle - \epsilon \label{eq:82}   \\
      &\geq -2 \delta \epsilon + \kappa - \epsilon = - \epsilon(2\delta + 1)+\kappa = 0, \label{eq:83}
    \end{align}
    where we have used Equation (\ref{eq:79}) in (\ref{eq:81}), Equation (\ref{eq:80}) in (\ref{eq:82}) and
    the definition of $\delta$ (\ref{eq:70}) in (\ref{eq:83}). 
  \end{pf}

  Now let us apply this lemma to $\rho_N^{(1)} = \rho$ and $\rho_N^{(2)} = \rho_N$ for all $N \in \Bbb{N}$. For each $\lambda \in
  (0,1)$ we get an $N_\lambda \in \Bbb{N}$ such that $ h_N(\rho,f) = h_N(\lambda \rho_N + (1-\lambda) \sigma_N,f)$ holds for all $N >
  N_\lambda$. Hence 
  \begin{equation} \label{eq:86}
    \lim_{N\to\infty} h_N(\rho,f) = \lim_{N \to \infty} h_N(\lambda \rho_N + (1-\lambda) \sigma_N,f).
  \end{equation}
  Using the definition of $h_N$ in (\ref{eq:84}) we get:
  \begin{multline}
    h_N(\lambda \rho_N + (1-\lambda) \sigma_N,f) = \frac{-1}{N} \ln \left( \lambda^N e^{-N h_N(\rho_N,f)} + \phantom{\sum_{n=1}^N \int_{\scr{S}}} \right. \\
      \left. \sum_{n=1}^N \lambda^{N-n} (1-\lambda)^n \int_{\scr{S}} e^{-N f(\sigma)} \tr\bigl(A_{N,n} E_N(d\sigma) \bigr) \right),
  \end{multline}
  where $A_{N,n}$ denotes the sum of all tensor products consisting of $N-n$ factors $\rho_N$ and $n$ factors
  $\sigma_N$. We can rewrite this expression as
  \begin{multline} \label{eq:85}
    h_N(\lambda \rho_N + (1-\lambda) \sigma_N,f) = -\ln \lambda + h_N(\rho_N,f) \\ 
    - \frac{1}{N} \ln \left( 1 + e^{N h_N(\rho_N,f)} \sum_{n=1}^N \left(\frac{1-\lambda}{\lambda}\right)^n \int_{\scr{S}} e^{-N f(\sigma)}
      \tr\bigl(A_{N,n} E_N(d\sigma) \bigr) \right).
  \end{multline}
  Since $\rho_N$ and $\sigma_N$ are density matrices, the operators $A_{N,n}$ are positive. Hence, the argument of the 
  last logarithm in Equation (\ref{eq:85}) is greater than one and the logarithm therefore positive. This
  implies:
  \begin{equation}
    h_N(\rho_N,f) \geq h_N(\lambda \rho_N + (1-\lambda) \sigma_N,f) + \ln(\lambda),
  \end{equation}
  and with Equation (\ref{eq:86})
  \begin{multline} \label{eq:88}
    \liminf_{N \to \infty} h_N(\rho_N,f) \geq \liminf_{N \to \infty} h_N(\lambda \rho_N + (1-\lambda) \sigma_N,f) + \ln(\lambda) = \\ \lim_{N \to \infty} h_N(\rho,f) +
    \ln(\lambda). 
  \end{multline}
  Since $\lambda \in(0,1)$ is arbitrary we get $\liminf_{N \to \infty} h_N(\rho_N,f) \geq \lim_{N \to \infty} h_N(\rho,f)$. The other
  inequality (i.e. $\limsup_{N \to \infty} h_N(\rho_N,f) \leq \lim_{N \to \infty} h_N(\rho,f)$) can be derived with the same
  argument, if we exchange the role of $\rho$ and $\rho_N$ (i.e. apply Lemma \ref{lem:3} to $\rho_N^{(1)} = \rho_N$ and
  $\rho_N^{(2)} = \rho$ for all $N \in \Bbb{N}$). Hence  $\lim_{N \to \infty} h_N(\rho_N,f) = \lim_{N \to \infty} h_N(\rho,f)$ as
  stated. The equality $\lim_{N \to \infty} h_N(\rho,f) = h(\rho,f)$ follows from Varadhan's Theorem (Theorem
  \ref{thm:7}). 

  Now consider statement \ref{item:3}. If $\rho$ is degenerate, the method used above can not be
  applied. However, if the rate function $I$ is sufficiently continuous, we can extend (parts of) the result
  derived for non-degenerate density matrices to the degenerate case. To this end we need the following lemma:

  \begin{lem} \label{lem:6}
    Consider a compact metric space $(X,d)$ and a lower semicontinuous function $F: X \times X \to [c,\infty]$, $c \in 
    \Bbb{R}$. The infimum $\underline{F}(x) = \inf_{y \in X} F(x,y)$ is lower semicontinuous as well.
  \end{lem}

  \begin{pf}
    Due to lower semicontinuity of $F$, we find for each $(x,y) \in X \times X$ and each $\epsilon > 0$ a $\delta_{x,y} > 0$
    with 
    \begin{equation}
      d(x,x') < \delta_{x,y},\ d(y,y') < \delta_{x,y} \Rightarrow F(x',y') > F(x,y) - \epsilon.
    \end{equation}
    Since $X$ is compact, each fixed $x \in X$ admits finitely many points $y_1, \ldots, y_k \in X$ such that the
    neighborhoods $U_j = \{ y' \in X \, | \, d(y',y_j) < \delta_{x,y_j}\}$ overlap $X$. Now define $\delta = \min_j
    \delta_{x,y_j} > 0$. For each $x'$ satisfying $d(x,x') < \delta$ and each $y' \in X$ there is a $j=1,\ldots,k$ 
    with $F(x',y') > F(x,y_j) - \epsilon$. Hence $F(x',y') > \inf_y F(x,y) - \epsilon$ and we get
    \begin{equation}
      d(x,x') < \delta \Rightarrow \underline{F}(x') = \inf_{y'} F(x',y') > \inf_y F(x,y) - \epsilon. 
    \end{equation}
    Since $\delta > 0$ this shows that $\underline{F}$ is lower semicontinuous at $x$ and since $x$ is arbitrary
    the statement follows.
  \end{pf}

  Let us apply this lemma to $F(\rho,\sigma) = I(\rho,\sigma) + f(\sigma)$. Since $I$ is lower semicontinuous by assumption we get
  for each $\epsilon > 0$ a $\delta > 0$ such that $\|\rho' - \rho\|_1 < \delta$ implies $h(\rho',f) > h(\rho,f) - \epsilon$. Together with the
  convexity of the $\delta$-ball around $\rho$ this implies 
  \begin{equation}
    h(\lambda\rho + (1-\lambda)\rho',f) > h(\rho,f) - \epsilon\quad \forall \lambda \in (0,1).
  \end{equation}
  If $(\rho_N)_{N \in \Bbb{N}}$ is a sequence in $\scr{S}$ converging to $\rho$, the convex linear combinations $\lambda \rho_N
  + (1-\lambda) \rho'$ converges to $\lambda \rho + (1-\lambda) \rho'$. As in Equation (\ref{eq:88}) we get 
  \begin{equation}
    \liminf_{N \to \infty} h_N(\lambda \rho_N + (1-\lambda) \rho',f) \leq \liminf_{N \to \infty}h_N(\rho_N,f) - \ln(\lambda).
  \end{equation}
  Now assume without loss of generality that $\rho'$ is non-degenerate. Then $\lambda \rho + (1-\lambda) \rho'$ is non-degenerate
  as well and we have according to item \ref{item:2} 
  \begin{equation}
    \liminf_{N \to \infty} h_N(\lambda \rho_N + (1-\lambda) \rho',f) = h(\lambda \rho + (1-\lambda) \rho',f) > h(\rho,f) - \epsilon.
  \end{equation}
  Hence
  \begin{equation}
     \liminf_{N \to \infty}h_N(\rho_N,f) \geq  h(\rho,f) - \epsilon + \ln(\lambda).
  \end{equation}
  Since $\epsilon > 0$ and $\lambda \in (0,1)$ are arbitrary the statement follows.
\end{pf}

According to Proposition 1.2.7 of \cite{DupEll} this lemma implies immediately that the convergence $h_N \to h$
is uniform on each compact set of non-degenerate density matrices.

\begin{prop}
  Consider the same assumptions as in the preceding lemma and a compact set $K \subset \scr{S}$ consisting only of
  non-degenerate density matrices. Then the convergence $h_N \to h$ is uniform on $K$, i.e. 
  \begin{equation}
    \lim_{N \to \infty} sup_{\rho \in K} \left| h_N(\rho,f) - h(\rho,f) \right| = 0 
  \end{equation}
  holds.
\end{prop}

Another simple consequence of Lemma \ref{lem:4} is the continuity of $h(\,\cdot\,,f)$ on the interior of
$\scr{S}$. The proof is again omitted, since it can be taken without change from \cite{DupEll} (first
paragraph of the proof of Proposition 1.2.7). 

\begin{prop} \label{prop:4}
  Consider again the assumptions from Lemma \ref{lem:4}. The function $\scr{S} \ni \rho \mapsto h(\rho,f) \in \Bbb{R}$ is
  continuous at each non-degenerate $\rho$.
\end{prop}

This is a somewhat surprising result, because it is derived without any further assumption on the rate
function $I$. Although it does not imply that $I(\rho,\sigma)$ is continuous in $\rho$, it shows at least that the
dependence of $I$ on the original density matrix $\rho$ is quite regular \emph{on the interior} of the state
space $\scr{S}$. On the boundary, however, nothing can be said. The discussion in Sections
\ref{sec:pure-states} and \ref{sec:beyond-covariance} will indicate that this is probably a fundamental aspect
of admissible rate functions and not just a problem of the methods used in the proofs.

Let us consider now the natural action of $\U(d)$ on the set $\scr{C}(\scr{S})$ of continuous functions on
$\scr{S}$, i.e. for each $U \in \U(d)$ and each $f \in \scr{C}(\scr{S})$ define $\alpha_Uf \in \scr{C}(\scr{S})$ by 
\begin{equation} \label{eq:99}
  \alpha_Uf(\sigma) = f(U\sigma U^*).
\end{equation}
Then we can consider for each fixed $\rho \in \scr{S}$ and each $f$ the functions
\begin{equation}
  \U(d) \ni U \mapsto h_N(U^*\rho U, \alpha_Uf) \in \Bbb{R}\ \text{and}\ \U(d) \ni U \mapsto h(U^*\rho U, \alpha_Uf) \in \Bbb{R},
\end{equation}
and pose the same question as above -- but now considering the dependency on $U$ rather than on $\rho$. The
following is the analog of Lemma \ref{lem:4}.

\begin{lem} \label{lem:5}
  Consider an estimation scheme $(E_N)_{N \in \Bbb{N}}$ satisfying the LDP with rate function $I$, an
  arbitrary continuous (real valued) function $f$ and the functionals $h_N, h$ defined in Equations
  (\ref{eq:84}) and (\ref{eq:97}).
  \begin{enumerate}
  \item \label{item:4}
    For each non-degenerate density matrix $\rho \in \scr{S}$ and each sequence $\Bbb{N} \ni N \mapsto U_N \in \U(d)$
    converging to $U \in \U(d)$ we have
    \begin{equation} \label{eq:102}
      \lim_{N \to \infty} h_N(U_N^*\rho U_N,\alpha_{U_N}f) = \lim_{N \to \infty} h_N(U^*\rho U,\alpha_Uf) = h(U^*\rho U,\alpha_Uf) 
    \end{equation}
  \item \label{item:5}
    If $I$ is lower semicontinuous \emph{in both arguments}, the lower bound
    \begin{equation}
      \liminf_{N \to \infty} h_N(U_N^*\rho U_N,\alpha_{U_N}f) \geq h(U^*\rho U,\alpha_Uf) 
    \end{equation}
    holds even for degenerate $\rho$.
  \end{enumerate}
\end{lem}

\begin{pf}
  To prove item \ref{item:4} let us start with the observation that the function sequence $(\alpha_{U_M}f)_{M \in
    \Bbb{N}}$ converges uniformly to $\alpha_Uf$: Due to compactness of $\scr{S}$ the function 
  $f$ is not just continuous but even uniformly continuous, i.e. for each $\epsilon > 0$ there is a $\delta > 0$ with
  \begin{equation} \label{eq:94}
    \|\sigma_1 - \sigma_2\|_1 < \delta \Rightarrow |f(\sigma_1) - f(\sigma_2)| < \epsilon.
  \end{equation}
  Convergence of $(U_M)_{M \in \Bbb{N}}$ implies the existence of $M_\epsilon \in \Bbb{N}$ with $M> M_\epsilon$ $\Rightarrow$ $\|U_M - U\| <
  \delta/2$. For each $\sigma$ and each $M > M_\epsilon$ we therefore get 
  \begin{align}
    \|U_M\sigma U_M^* - U\sigma U^*\|_1 &\leq \|U_M\sigma U_M^* - U_M\sigma U^*\|_1 + \| U_M\sigma U^* - U\sigma U^*\|_1 \\
    &\leq \|U_M^* - U^*\| \|U_M\| \|\sigma\|_1 + \|U_M - U\|  \|U^*\| \|\sigma\|_1 < \delta,
  \end{align}
  which implies together with (\ref{eq:94}) for an \emph{arbitrary} $\sigma$ and $M > M_\epsilon$
  \begin{equation}
    | \alpha_{U_M}f(\sigma) - \alpha_Uf(\sigma) | = |f(U_M\sigma U_M^*) - f(U\sigma U^*)| < \epsilon.
  \end{equation}
  In other words the convergence $\alpha_{U_M}f \to \alpha_Uf$ is uniform as stated (since $M_\epsilon$ does not depend on $\sigma$).

  To proceed, it is necessary to consider the following simple properties of the functionals $h_N$ and $h$: If
  $f, f_1$ denotes continuous functions on $\scr{S}$ and $\epsilon \in \Bbb{R}$ we have for all $\rho$
  \begin{equation} \label{eq:95}
    f \geq f_1 \Rightarrow h_N(\rho,f) \geq h_N(\rho,f_1),\ \text{and}\ h_N(\rho,f+\epsilon) = h_N(\rho,f)+\epsilon,
  \end{equation}
  and from Lemma \ref{lem:4} we already know that for all $\epsilon>0$ and all $f$ there is an $N[\epsilon,f] \in \Bbb{N}$ with 
  \begin{equation} \label{eq:96}
    N > N[\epsilon,f] \Rightarrow | h_N(U_N^*\rho U_N,f) - h(U^*\rho U,f) | < \epsilon.
  \end{equation}
  Uniform convergence $\alpha_{U_M}f \to \alpha_Uf$ implies that $\alpha_Uf - \epsilon \leq \alpha_{U_M}f \leq \alpha_Uf + \epsilon$ holds for all $M >
  M_\epsilon$. Hence for all $N \in \Bbb{N}$ we have
  \begin{equation} \label{eq:98}
    h_N(U_N^*\rho U_N, \alpha_Uf) - \epsilon \leq  h_N(U_N^*\rho U_N, \alpha_{U_M}f) \leq h_N(U_N^*\rho U_N, \alpha_Uf) + \epsilon
  \end{equation}
  according to (\ref{eq:95}). Together with (\ref{eq:96}) we get
  \begin{equation}
    N > N[\epsilon,\alpha_Uf],\ M > M_\epsilon \Rightarrow | h_N(U_N^*\rho U_N,\alpha_{U_M}f) - h(U^*\rho U, \alpha_Uf)| < 2\epsilon,
  \end{equation}
  which implies Equation (\ref{eq:102}).

  Statement \ref{item:5} can be shown in the same way, if we replace Equation (\ref{eq:96}) by (cf. Lemma
  \ref{lem:4}) 
  \begin{equation} 
    N > N[\epsilon,f] \Rightarrow  h_N(U_N^*\rho U_N,f) \geq h(U^*\rho U,f) - \epsilon
  \end{equation}
  and use only the lower bound of (\ref{eq:98}).
\end{pf}

As in the case of Lemma \ref{lem:4} we can now derive continuity and uniformity properties from this
result. The following proposition is (again) an immediate consequence of \cite[Prop. 1.2.7]{DupEll}. The
proof is therefore omitted.

\begin{prop} \label{prop:5}
  Consider the same assumptions as in Lemma \ref{lem:5} and a non-degenerate density matrix $\rho$.
  \begin{enumerate}
  \item 
    The function  
    \begin{equation}
      \U(d) \ni U \mapsto h(U^*\rho U,\alpha_Uf) = \inf_{\sigma \in \scr{S}} \bigl( I(U^*\rho U, U^*\sigma U) + f(\sigma) \bigr) \in \Bbb{R}
    \end{equation}
    is continuous.
  \item 
    The convergence of $h_N(U^*\rho U, \alpha_Uf)$ to $h(U^*\rho U, \alpha_Uf)$ is uniform in $U$, i.e.
    \begin{equation}
      \lim_{ N \to \infty} \sup_{U \in \U(d)} | h_N(U^*\rho U, \alpha_Uf) - h(U^*\rho U, \alpha_Uf)| = 0
    \end{equation}
    holds.
  \end{enumerate}
\end{prop}

%%%%%%%%%%%%%%%%%%%%%%%%%%%%%%%%%%%%%%%%%%%%%%%%%%
\subsection{Averaging}
\label{sec:averaging}
%%%%%%%%%%%%%%%%%%%%%%%%%%%%%%%%%%%%%%%%%%%%%%%%%%

Let us consider now the question whether covariance and permutation invariance are ``harmful'' for the rate
function; i.e. can we hope to exhaust the optimal upper bounds from Equation (\ref{eq:22}) with schemes
admitting these symmetry properties? One possible way to answer this question is to start with a general
scheme $(E_N)_{N \in   \Bbb{N}}$ and to \emph{average} over the unitary and the permutation group. For the
latter this leads to   
\begin{equation} 
  \overline{E}_N(\Delta) = \frac{1}{N!} \sum_{p \in \Sym_N} V_p E_N(\Delta) V_p^*,
\end{equation} 
and since we have
\begin{equation}
  \tr \bigl(\rho^{\otimes N} V_p E_N(\Delta) V_p^* \bigr) =  \tr \bigl(V_p^* \rho^{\otimes N} V_p E_N(\Delta)\bigr)
  =  \tr \bigl(\rho^{\otimes N} E_N(\Delta)\bigr)
\end{equation}
for each permutation $p \in \Sym_N$, we see that the rate function is not changed at all by this procedure. Hence,
for the rest of this section we can assume without loss of generality that each scheme is permutation
invariant.

This leads us to averages over the unitary group, i.e.
\begin{equation} \label{eq:38}
  \overline{E}_N(\Delta) = \int_{\U(d)} U^{\otimes N} E_N(U^* \Delta U) U^{\otimes N*} dU.
\end{equation}
Here the situation is (unfortunately) different. The following proposition shows that the convergence behavior
of $\overline{E}_N$ is in general worse than that of $E_N$. 

\begin{prop} \label{prop:6}
  Consider an estimation scheme $(E_N)_{N \in \Bbb{N}}$ satisfying the LDP with rate function $I$ and the
  corresponding averaged scheme $(\overline{E}_N)_{N \in \Bbb{N}}$ from Equation (\ref{eq:38}). For each
  non-degenerate density matrix $\rho$ the sequence of probability measures $\tr\bigl(\rho^{\otimes N}
  \overline{E}_N(\,\cdot\,) \bigr)$ satisfies the LDP with rate function $\overline{I}_\rho$ given by
  \begin{equation}
    \overline{I}_\rho(\sigma) = \overline{I}(\rho,\sigma) = \inf_{U \in \U(d)} I(U^*\rho U, U^*\sigma U).
  \end{equation}
\end{prop}

\begin{pf}
  It is sufficient to show that the measures $\tr\bigl(\rho^{\otimes N} \overline{E}_N(\,\cdot\,)\bigr)$ satisfy the
  \emph{Laplace principle} with the same rate function  (cf. Theorem \ref{thm:8}), because the Laplace
  principle is equivalent to the large deviation principle. Hence we have to show that
  \begin{equation} \label{eq:40}
    \lim_{N \to \infty} \frac{-1}{N} \ln \int_\scr{S} e^{-N f(\sigma)} \tr \bigl( \rho^{\otimes N} \overline{E}_N(d\sigma) \bigr) = 
    \inf_{\sigma \in \scr{S}} \bigl(f(\sigma) + \overline{I}(\rho,\sigma)\bigr)
  \end{equation}
  holds for all continuous functions $f$ on $\scr{S}$. Inserting the definition of $\overline{E}_N$ we get 
  \begin{multline} \label{eq:42}
    \int_\scr{S} e^{-N f(\sigma)} \tr \bigl( \rho^{\otimes N} \overline{E}_N(d\sigma) \bigr) \\ =  \int_{\U(d)} \int_\scr{S} e^{-N f(U \sigma U^*)} \tr
    \bigl( (U^*\rho U)^{\otimes N} E_N(d\sigma) \bigr) dU,
  \end{multline}
  or with the notation from Subsection \ref{sec:cont-prop} (cf. Equations (\ref{eq:84}) and (\ref{eq:99}))
  \begin{equation} \label{eq:100}
    \int_\scr{S} e^{-N f(\sigma)} \tr \bigl( \rho^{\otimes N} \overline{E}_N(d\sigma) \bigr) = \int_{\U(d)} e^{-N h_N(U^*\rho U,\alpha_Uf)} dU.
  \end{equation}
  According to Proposition \ref{prop:5} the quantity $h_N(U^*\rho U,\alpha_Uf)$ converges \emph{uniformly in} $U$ to 
  $h(U^*\rho U,\alpha_Uf)$, i.e. for each $\epsilon > 0$ there is an $N_\epsilon \in \Bbb{N}$ such that
  \begin{multline} 
    N > N_\epsilon \Rightarrow \\  h(U^*\rho U,\alpha_Uf) + \epsilon \geq h_N(U^*\rho U,\alpha_Uf) \geq h(U^*\rho U,\alpha_Uf)  - \epsilon\quad \forall U \in \U(d)
  \end{multline}
  holds. Hence, for each $\epsilon > 0$ we get
  \begin{multline}
    \limsup_{N \to \infty} \frac{-1}{N} \ln \int_{\U(d)} e^{-N  (h(U^*\rho U,\alpha_Uf) + \epsilon)} dU \geq \\ 
    \limsup_{N \to \infty} \frac{-1}{N} \ln \int_{\U(d)} e^{-N h_N(U^*\rho U,\alpha_Uf)} dU. 
  \end{multline}
  From Proposition \ref{prop:5} we know that $h(U^*\rho U,\alpha_Uf)$ is continuous in $U$ and we can apply
  Varadhan's Theorem (Theorem \ref{thm:7}) to the left hand side of this inequality. Together with
  \begin{align}
    \inf_{U \in \U(d)} h(U^*\rho U,\alpha_Uf) & = \inf_{U \in \U(d)} \inf_{\sigma \in \scr{S}} \bigl( I(U^*\rho U,\sigma) + f(U\sigma U^*)\bigr) \\
    &= \inf_{U \in \U(d)} \inf_{\sigma \in \scr{S}} \bigl( I(U^*\rho U,U^*\sigma U) + f(\sigma)\bigr) \\
    &= \inf_{\sigma \in \scr{S}} \bigl(\overline{I}(\rho,\sigma) + f(\sigma)\bigr)
  \end{align}
  this implies the upper bound
  \begin{equation} \label{eq:101}
    \limsup_{N \to \infty} \frac{-1}{N} \ln \int_{\U(d)} e^{-N h_N(U^*\rho U,\alpha_Uf)} dU \leq \inf_{\sigma \in \scr{S}}
    \bigl(\overline{I}(\rho,\sigma) + f(\sigma)\bigr) + \epsilon. 
  \end{equation}
  The lower bound 
    \begin{equation} \label{eq:103}
    \liminf_{N \to \infty} \frac{-1}{N} \ln \int_{\U(d)} e^{-N h_N(U^*\rho U,\alpha_Uf)} dU \geq \inf_{\sigma \in \scr{S}}
    \bigl(\overline{I}(\rho,\sigma) + f(\sigma)\bigr) - \epsilon 
  \end{equation}
  can be shown in the same way. Since $\epsilon > 0$ was arbitrary, Equation (\ref{eq:40}) follows from 
  (\ref{eq:100}), (\ref{eq:101}) and (\ref{eq:103}), which concludes the proof.
\end{pf}

Hence, the best we can hope is that the averaged scheme satisfies the LDP with rate function $\overline{I}$
which is actually the worst $\U(d)$-invariant rate function which can be derived from $I$. Only if $I$ is
$\U(d)$ invariant itself (such that $\overline{I} = I$ holds), the convergence behavior of
$(\overline{E}_N)_{N \in \Bbb{N}}$ is as good as than that of $(E_N)_{N \in \Bbb{N}}$. The following proposition
shows that at least in this case the convergence problems on the boundary of $\scr{S}$ can be solved.

\begin{prop} \label{prop:1}
  If $(E_N)_{N \in \Bbb{N}}$ is an estimation scheme satisfying the LDP with a $\U(d)$-invariant, lower 
  semicontinuous (in both arguments) rate function $I$, the averaged scheme $(\overline{E}_N)_{N \in \Bbb{N}}$
  defined in Equation (\ref{eq:38}) satisfies the LDP with the same rate function. 
\end{prop}

\begin{pf}
  We will show again the alternative statement that the sequence $\tr\bigl(\rho^{\otimes N}
  \overline{E}_N(\,\cdot\,)\bigr)$ satisfies the Laplace principle, i.e. Equation (\ref{eq:40}) holds for all
  continuous real valued functions $f$ and with $\overline{I}$ replaced by $I$. As in the last proof we can
  rewrite this in terms of the functionals $h_N$ and $h$ defined in Equation (\ref{eq:84}) and (\ref{eq:97}),
  i.e. we have to show that (cf. Equation (\ref{eq:100}))
  \begin{equation} \label{eq:108}
    \limsup_{N \to \infty} \frac{-1}{N} \ln \int_{\U(d)} e^{-N h_N(U^*\rho U, \alpha_Uf)} dU \leq h(\rho,f) 
  \end{equation}
  and
  \begin{equation} \label{eq:109}
    \liminf_{N \to \infty} \frac{-1}{N} \ln \int_{\U(d)} e^{-N h_N(U^*\rho U, \alpha_Uf)} dU \geq h(\rho,f) 
  \end{equation}
  hold. But now the convergence of $h_N(U^*\rho U, \alpha_Uf)$  to $h(\rho,f)$ is only known to be pointwise (and not
  necessarily uniform) in $U$. Therefore, we can not proceed as in Proposition \ref{prop:6}. Instead we will
  use different strategies for the upper and the lower bound. 

  To get the upper bound note that $f$ is (as a continuous function on a compact set) bounded from above by a
  constant $K > 0$. Therefore the functions $U \mapsto h_N(U^*\rho U,\alpha_Uf)$ are bounded as well (by the same
  constant) and we get (note that $h(U^*\rho U,\alpha_Uf) = h(\rho,f)$ holds for all $U$ by assumption)
  \begin{equation} \label{eq:104}
    \lim_{N \to \infty} \int_{\U(d)} \bigl| h_N(U^*\rho U, \alpha_Uf) - h(\rho,f) \bigr| dU = 0.
  \end{equation}
  by the dominated convergence theorem. Now let us introduce for each $\epsilon > 0$ and each $N \in \Bbb{N}$ the 
  set
  \begin{equation}
    \Delta_{N,\epsilon} = \{ U \in \U(d)\,|\, \bigl| h_N(U^*\rho U, \alpha_Uf) - h(\rho,f) \bigr| > \epsilon \}. 
  \end{equation}
  From Equation (\ref{eq:104}) we see that for each $\delta > 0$ there is an $N_\delta \in \Bbb{N}$ such that $N > N_\delta$
  implies
  \begin{equation}
    \left| \Delta_{N,\epsilon} \right| \epsilon \leq \int_{\U(d)}  \bigl| h_N(U^*\rho U, \alpha_Uf) - h(\rho,f) \bigr| dU < \delta,
  \end{equation}
  where $\left| \Delta_{N,\epsilon} \right|$ denotes the volume of $\Delta_{N,\epsilon}$ with respect to the Haar measure (note that
  $\Delta_{N,\epsilon}$ is due to continuity of $U \mapsto h_N(U^*\rho U,\alpha_Uf)$ open and therefore measurable). Now choose
  $\epsilon>0$ arbitrary and $\delta=\epsilon/2$ then we have for all $N > N_\delta$
  \begin{equation}
    \int_{\U(d)} e^{-N h_N(U^*\rho U, \alpha_Uf)} dU \geq \int_{\U(d) \setminus \Delta_{N,\epsilon}} \hspace{-5.2ex} e^{-N h_N(U^*\rho U, \alpha_Uf)} dU \geq
    \frac{1}{2} e^{-N (h(\rho,f) + \epsilon)},  
  \end{equation}
  where we have used the fact that $h_N(U^*\rho U, \alpha_Uf) < h(\rho,f) + \epsilon$ holds for all $U \not\in \Delta_{N,\epsilon}$. Taking
  logarithms and the limit $N \to \infty$ this implies
  \begin{equation}
    \limsup_{N \to \infty} \frac{-1}{N} \ln \int_{\U(d)} e^{-N h_N(U^*\rho U, \alpha_Uf)} dU \leq h(\rho,f) + \epsilon.
  \end{equation}
  Since $\epsilon>0$ was arbitrary we get the upper bound (\ref{eq:108}).

  To prove the lower bound let us assume first that
  \begin{equation} \label{eq:107}
    \liminf_{N \to \infty} \inf_{U \in \U(d)} \bigl( h_N(U^*\rho U,\alpha_Uf) - h(\rho,f) \bigr) \geq 0
  \end{equation}
  does not hold. Then we can find a sequence $(U_N)_{N \in \Bbb{N}}$ of unitaries with
  \begin{equation} \label{eq:105}
    \liminf_{N \to \infty}  \bigl( h_N(U_N^*\rho U_N,\alpha_{U_N}f) - h(\rho,f) \bigr) < 0. 
  \end{equation}
  But due to compactness of $\U(d)$ we can assume without loss of generality that $(U_N)_{N \in \Bbb{N}}$
  converges to a unitary $U$. Hence Equation (\ref{eq:105}) contradicts Statement \ref{item:5} of Lemma
  \ref{lem:5} (since the rate function $I$ is lower semicontinuous by assumption).  Hence Equation
  (\ref{eq:107}) is valid and we can find for each $\epsilon>0$ an $N_\epsilon \in \Bbb{N}$ such that $N > N_\epsilon$ implies
  \begin{equation}
    h_N(U^*\rho U,\alpha_{U}f) >  h(\rho,f) - \epsilon\quad \forall U \in \U(d).
  \end{equation}
  Hence
  \begin{equation}
    \liminf_{N \to \infty} \frac{-1}{N} \ln \int_{\U(d)} e^{-N h_N(U^*\rho U,\alpha_{U}f)} dU > h(\rho,f) - \epsilon.
  \end{equation}
  Since $\epsilon>0$ is arbitrary we get the lower bound (\ref{eq:109}) and the proof is completed.
\end{pf}

This result is very useful if we want to check whether a given rate function is admissible or not. Many
prominent candidates are $\U(d)$-invariant and lower semicontinuous (like relative entropy), and in this case
it is according to Proposition \ref{prop:1} sufficient to consider only covariant schemes. Important examples
of functions which can be tested this way are the optimal rate functions $\scr{I}_{\Id}$ and $\scr{I}^0_{\Id}$ (for
$\scr{I}_{\Id}$ this is true at least on the interior of $\scr{S}$): 

\begin{prop} \label{prop:3}
  The optimal rate functions $\scr{I}_{\Id}$ and $\scr{I}_{\Id}^0$ are $\U(d)$ invariant (i.e. Equation (\ref{eq:48})
  holds with $\alpha_U(\sigma) = U\sigma U^*$). 
\end{prop}

\begin{pf}
  Since $\scr{I}_{\Id}$ and $\scr{I}_{\Id}^0$ are defined as the upper bounds on $\scr{E}(\Id)$ and $\scr{E}^0(\Id)$
  we have to show that these sets are invariant under the operation $I \mapsto I_U$ with $I_U(\rho,\sigma) =
  I(U\rho U^*,U\sigma U^*)$. Hence consider $I \in \scr{E}(\Id)$. Then there is a full estimation scheme $(E_N)_{N \in
    \Bbb{N}}$ satisfying LDP with rate function $I$. For each fixed $U \in \U(d)$ we can define the translated
  scheme $(E_N^U)_{N \in \Bbb{N}}$ with $E_N^U(\Delta) = U^{\otimes N *} E_N(U \Delta U^*) U^{\otimes N}$. If $\Delta$ is open we get
  \begin{align}
    \liminf_{N \to \infty} \frac{1}{N} \ln \tr \bigl( \rho^{\otimes N} E_N^U(\Delta) \bigr) &= \liminf_{N \to \infty} \frac{-1}{N} \ln \tr
    \bigl( (U\rho U^*)^{\otimes N} E_N(U\Delta U^*) \bigr) \\
    & \leq - \inf_{\sigma \in U\Delta U^*} I(U\rho U^*,\sigma) \\ & = - \inf_{\sigma \in \Delta} I(U\rho U^*,U\sigma U^*).
  \end{align}
  This shows that the large deviation upper bound holds with rate function $I_U$. The lower bound can be shown
  in the same way. Hence $(E_N^U)_{N \in \Bbb{N}}$ satisfies the LDP with rate function $I_U$, and this implies
  $I_U \in \scr{E}(\Id)$. Since the operation $I \mapsto I_U$ respects semi-continuity of $I$, 
  invariance of $\scr{E}^0(\Id)$ is trivial and this concludes the proof.  
\end{pf}

Summarizing the discussion of this subsection we can conclude that averaging is in the context of large deviations
not as powerful as it is in other areas like optimal cloning. Nevertheless, it is not completely useless either. In
particular the conjecture $\scr{I}_{\Id}^0 \in \scr{E}(p)$ is interesting in this regard, because it would imply
that $\scr{I}_{\Id}^0$ can be derived as the rate function of a covariant scheme. Hence, covariant schemes are
an important special case (and therefore worth studying), although they probably can not tell us the whole
truth. 

%%%%%%%%%%%%%%%%%%%%%%%%%%%%%%%%%%%%%%%%%%%%%%%%%%
\subsection{General structure}
\label{sec:general-structure}
%%%%%%%%%%%%%%%%%%%%%%%%%%%%%%%%%%%%%%%%%%%%%%%%%%

Now let us have a look at the general structure of covariant and permutation invariant estimation schemes. Our
main tool is the following theorem about covariant observables \cite{HolBook}.

\begin{thm} \label{thm:6}
  Consider a compact group $G$ which acts transitively on a locally compact, separable metric space $X$ by
  $G \times X \ni (g,x) \mapsto \alpha_g(x)$, and a representation $\pi$ of $G$ on a Hilbert space $\scr{H}$. Each POV measure $E:
  \goth{B}(X) \to \scr{B}(\scr{H})$ which is \emph{covariant} (i.e. $E(\alpha_g\Delta) = \pi(g) E(\Delta) \pi(g)^*$ for all $\Delta \in 
  \goth{B}(X)$ and all $g \in G$) has the form 
  \begin{equation} \label{eq:106}
    \int_X f(x) E(dx) = \int_G f(\alpha_gx_0) \pi(g) Q_0 \pi(g)^* \mu(dg) 
  \end{equation}
  where $x_0 \in X$ is an (arbitrary) reference point, $\mu$ is the Haar-measure on $G$ and $Q_0 \in
  \scr{B}(\scr{H})$ a positive operator which is uniquely determined by (\ref{eq:106}) and the choice of
  $x_0$. 
\end{thm}

Unfortunately this theorem is not applicable to our case, because the action of $\U(d)$ on $\scr{S}$ is
not transitive. A way out of this dilemma is to look at the fibration $s: \scr{S} \to \Sigma$ defined in Equation
(\ref{eq:21}) and to apply the results about transitive group actions to each fiber separately. (For the rest
of this section we will use frequently the notations introduced in Section \ref{sec:estimating-spectrum-1}.)

\begin{thm} \label{prop:2}
  Each covariant and permutation invariant observable $E: \goth{B}(\scr{S}) \to \scr{B}(\scr{H}^{\otimes N})$ has the
  form (with a continuous function $f$ on $\scr{S}$)
  \begin{multline} \label{eq:32}
    \int_\scr{S} f(\rho) E(d\rho) = \\ \sum_{Y \in \scr{Y}_d(N)} \left[ \int_{U(d)} \pi_Y(U) \left( \int_\Sigma f(U\rho_xU^*) q_Y(dx)
      \right) \pi_Y(U^*) dU \right] \otimes \Bbb{1}_Y
  \end{multline}
  with a sequence of (non-normalized) POV measures $q_Y : \goth{B}(\Sigma) \to \scr{B}(\scr{H}_Y)$, the diagonal
  matrices $\rho_x = \diag(x_1,\ldots,x_d)$ from Equation (\ref{eq:9}) and the unit matrix $\Bbb{1}_Y \in
  \scr{B}(\scr{K}_Y)$.
\end{thm}

\begin{pf}
  Permutation invariance implies immediately that 
  \begin{equation} \label{eq:35}
    E_N(\Delta) = \bigoplus_{Y \in \scr{Y}_d(N)} E_{N,Y}(\Delta) \otimes \Bbb{1}_Y
  \end{equation}
  holds with $\Bbb{1}_Y \in \scr{B}(\scr{K}_Y)$ and a family of POV measures $E_{N,Y} : \goth{B}(\scr{S}) \to
  \scr{B}(\scr{H}_Y)$, which are again $\U(d)$ covariant:
  \begin{equation}
    E_{N,Y}(U\Delta U^*) =\pi_Y(U) E_N(\Delta)\pi_Y(U)^*\quad \forall U \in \U(d).
  \end{equation}
  Hence we only have to look at $E_{N,Y}$ for a fixed $Y \in \scr{Y}_d(N)$, Therefore the statement is a
  consequence of the following lemma. 

  \begin{lem}
    Each $\U(d)$ covariant observable $E: \goth{B}(\scr{S}) \to \scr{B}(\scr{H}_Y)$ has the form
    \begin{equation} \label{eq:33}
      \int_\scr{S} f(\rho) E(d\rho)  = \int_{\U(d)} \pi_Y(U) \left( \int_\Sigma f(U\rho_xU^*) q(dx) \right) \pi_Y(U^*) dU
    \end{equation}
    with an appropriate POV-measure $q: \goth{B}(\Sigma) \to \scr{B}(\scr{H}_Y)$.
  \end{lem}

  \begin{pf}
    To each $\rho \in G$ we can associate the stabilizer subgroup $G_\rho = \{ U \in \U(d)\, | \, U\rho U^* = \rho \}$ of
    $\U(d)$, whose structure is uniquely determined by the degeneracy of the eigenvalues of $\rho$. Hence 
    the set
    \begin{equation}
      J = \{ G_{\rho_x} \, | \, x \in \Sigma \}\ \text{with}\ \rho_x = \diag(x_1, \ldots, x_d)
    \end{equation}
    is finite and for each $\rho$ there is exactly one $G \in J$ such that $G_\rho = UGU^*$ holds with an appropriate
    unitary $U \in \U(d)$. We can decompose $\scr{S}$ therefore into a disjoint union $\scr{S} = \bigcup_{G \in J}
    \scr{S}_G$ of finitely many subsets\footnote{The decomposition of $\scr{S}$ into a finite union of fiber
      bundles we are describing here is a special case of a much more general result (``slice theorem'') about
      compact $G$-manifolds; cf. \cite{Jaenich68}.}  
    \begin{equation}
      \scr{S}_G = \{ \rho \in \scr{S}\, | \, \exists U \in \U(d) \ \text{with}\ G_\rho = UGU^* \};
    \end{equation}
    and similarly we have $\Sigma = \bigcup_G \Sigma_G$ with $\Sigma_G = s(\scr{S}_G)$. By construction each orbit $s^{-1}(x)$, $x
    \in \Sigma_G$ is naturally homeomorphic to the homogeneous space $X_G = \U(d) / G$. Hence, there is a natural
    homeomorphism $\Phi_G : \Sigma_G \times X_G \to \scr{S}_G$ which is uniquely determined by
    \begin{equation} \label{eq:62}
      \Phi_G(x, [\Bbb{1}]) = \rho_x\ \text{and}\ \Phi_G(x, [V]) = V\rho_xV^*\quad \forall x \in \Sigma_G\ \forall [V] \in X_G. 
    \end{equation}
    Note that the crucial property of $\Phi_G$ is to intertwine the group actions $\rho \mapsto U\rho U^*$ and $[V] \mapsto [UV]$
    of $\U(d)$ on $\scr{S}_G$ and $X_G$ respectively. 

    The $\scr{S}_G$ are in general neither open nor closed, but they are Borel subsets of $\scr{S}$ (more
    precisely differentiable submanifolds with boundary): Since $s$ is continuous, it is obviously sufficient
    to show that $\Sigma_G \in \goth{B}(\Sigma)$ holds. But this follows from the fact that each $\Sigma_G$ can be expressed as
    the complement of a Borel set in a finite union of closed sets (this is easy to see but tedious to write
    down). $\scr{S}_G \in \goth{B}(\scr{S})$ now implies $\goth{B}(\scr{S}_G) = \{ \Delta \cap \scr{S}_G  \, | \, \Delta \in
    \goth{B}(\scr{S}) \} \subset \goth{B}(\scr{S})$ and we can define the POV measures $E_G : \goth{B}(\scr{S}_G) \to
    \scr{B}(\scr{H}_Y)$, $E_G(\Delta) = E(\Delta)$. Note that the $E_G$ are not normalized and some of them can vanish
    completely. Since we can reconstruct $E$ from the $E_G$ by $E(\Delta) = \sum_G E_G(\Delta \cap \scr{S}_G)$ it is
    sufficient to prove the statement for each $G$ separately. In addition we can use the homeomorphism $\Phi_G$
    from Equation (\ref{eq:62}) to identify $\scr{S}_G$ with $\Sigma_G \times X_G$ and $E_G$ with a POVM on
    $\goth{B}(\Sigma_G \times X_G)$ which is covariant with respect to the group action 
    \begin{equation}
      \Sigma_g \times X_G \ni (x, [V]) \mapsto \alpha^G_U(x,[V]) = (x, [UV]) \in \Sigma_G \times X_G
    \end{equation}
    of $\U(d)$, i.e.
    \begin{equation} \label{eq:63}
      E_G(\alpha^G_U \Delta) = \pi_Y(U) E_G(\Delta) \pi_Y(U^*)\quad \forall \Delta \in \goth{B}(\Sigma_G \times X_G)\ \forall U \in \U(d).
    \end{equation}
    This is a direct consequence of the intertwining property of $\Phi_G$ mentioned above.

    Now let us consider the Abelian algebras $\scr{C}(X_G)$ and $\scr{C}(\Sigma_G)$ of continuous functions on $X_G$
    and $\Sigma_G$. Each $h \in \scr{C}(\Sigma_G)$ defines a positive linear map by
    \begin{equation} \label{eq:51}
      \scr{C}(X_G) \ni k \mapsto \tilde{E}_{G,h}(k) = \int_{\Sigma_G \times X_G} h(x) k(y) E_G(dx \times dy) \in \scr{B}(\scr{H}_Y).
    \end{equation}
    Positivity and linearity of $\tilde{E}_{G,h}$ imply that it can be expressed as an integral over $X_G$
    with respect to a POV measure $E_{G,h}$ 
    \begin{equation}
      \tilde{E}_{G,h}(k) = \int_{X_G} k(y) E_{G,h}(dy)
    \end{equation}
    (this is a general property of positive maps on Abelian algebras; cf. \cite{Paulsen}). From (\ref{eq:63})
    it follows immediately that $E_{G,h}$ is covariant and we  can apply Theorem \ref{thm:6}, i.e. there is a
    positive operator $Q_G(h)$ such that 
    \begin{equation}
      \tilde{E}_{G,h}(k) = \int_{\U(d)} k([U]) \pi_Y(U) Q_G(h) \pi_Y(U^*) dU 
    \end{equation}
    holds. Note that the distinguished point $x_o$ from Theorem \ref{thm:6} is in our case $[\Bbb{1}] \in
    X_G$. Since $Q_Y(h)$ is uniquely defined by this equation (cf. Theorem \ref{thm:6}) we get another
    positive linear map $Q_G: \scr{C}(\Sigma_G) \ni h \mapsto Q(h) \in \scr{B}(\scr{H}_Y)$ which can again be expressed as an
    integral 
    \begin{equation}
      Q_G(h) = \int_{\Sigma_G} h(x) q_G(dx),
    \end{equation}
    and we get 
    \begin{multline} \label{eq:64}
      \int_{\Sigma_G \times X_G} f(x, y) E_G(dx \times dy) = \\ \int_{\U(d)}  \pi_Y(U) \left( \int_{\Sigma_G} f([U],x) q_G(dx)
      \right) \pi_Y(U^*) dU.
    \end{multline}
    for each $f$ of the form $f(x,y) = k(x) h(y)$ with $k \in \scr{C}(\Sigma_G)$, $h \in \scr{C}(X_G)$, and by
    linearity and continuity for each continuous $f$ on $\Sigma_G \times X_G$. Now we can again apply the homeomorphism
    $\Phi_G$ to map $E_G$ back to a measure on $\scr{S}_G$. Since $\Phi_G$ intertwines the action of $\U(d)$ on
    $\scr{S}_G$ and $\Sigma_G  \times X_G$ we get from (\ref{eq:64}) 
    \begin{equation}
      \int_\scr{S} f(\rho) E_G(d\rho)  = \int_{\U(d)} \pi_Y(U) \left( \int_\Sigma f(U\rho_xU^*) q_G(dx) \right) \pi_Y(U^*) dU
    \end{equation}
    Hence the statement of the lemma follows with $q(\Delta) = \sum_G q_G(\Delta \cap \Sigma_G)$.
  \end{pf}

  Together with the decomposition of $E$ from Equation (\ref{eq:35}) the statement of this lemma concludes the
  proof of the theorem.
\end{pf}
  
%%%%%%%%%%%%%%%%%%%%%%%%%%%%%%%%%%%%%%%%%%%%%%%%%%
\subsection{An explicit scheme}
\label{sec:an-explicit-scheme}
%%%%%%%%%%%%%%%%%%%%%%%%%%%%%%%%%%%%%%%%%%%%%%%%%%

The class of observables described in Theorem \ref{prop:2} is still quite big. To reduce the freedom of
choice further we can focus our attention to estimation schemes which coincide with $(\hat{F}_N)_{N \in
  \Bbb{N}}$ from Theorem \ref{thm:5}, as long as only information about the spectrum of $\rho$ is required. In
other words $E_N$ should satisfy for all $N \in \Bbb{N}$ 
\begin{equation} \label{eq:44}
  E_N\bigl(s^{-1}(\Delta)\bigr) = \hat{F}_N(\Delta)\quad \forall \Delta \in \goth{B}(\Sigma),
\end{equation}
%where $s: \scr{S} \to \Sigma$ denotes again the projection from (\ref{eq:21}). 
This leads to the following corollary.

\begin{cor} \label{kor:1}
  Each covariant and permutation invariant estimation scheme $(E_N)_{N \in \Bbb{N}}$ which satisfies Equation
  (\ref{eq:44}) can be written as
  \begin{equation} \label{eq:31}
    \int_\scr{S} f(\rho) E_N(d\rho) = \sum_{Y \in \scr{Y}_d(N)} \int_{\U(d)} f(U \rho_{Y/N} U^*) U^{\otimes N} (Q_Y \otimes \Bbb{1}) U^{\otimes N *} dU,
  \end{equation}  
  with a family of operators $Q_Y \in \scr{B}(\scr{H}_Y)$.
\end{cor}

\begin{pf}
  Equation (\ref{eq:44}) implies immediately that the POV measures $q_Y$ from Proposition \ref{prop:2} are
  discrete, i.e. 
  \begin{equation}
    q_Y = \sum_{Z \in \scr{Y}_d(N)} q_{YZ} \delta_{Z/N}
  \end{equation}
  where $\delta_{Z/N}$ denotes the Dirac measure at $Z/N \in \Sigma$ and $q_{YZ} \in \scr{B}(\scr{H}_Y)$. Hence $E_N$
  becomes 
  \begin{equation}
    \int_\scr{S} f(\rho) E_N(d\rho) = \sum_{Y \in \scr{Y}_d(N)} \int_{\U(d)} f(U \rho_{Y/N} U^*) U^{\otimes N} \tilde{Q}_Y U^{\otimes N *} dU,
  \end{equation}  
  with 
  \begin{equation} \label{eq:50}
    \tilde{Q}_Y = \sum_{Z \in \scr{Y}_d(N)} q_{ZY} \otimes \Bbb{1}.
  \end{equation}
  Using the definition of $\hat{F}_N$ in Equation (\ref{eq:4}) and again Equation (\ref{eq:44}) we get
  \begin{equation}
    P_Y = \hat{F}_N\bigl(\{Y/N\}\bigr) = E_N\bigl(s^{-1}(Y/N)\bigr) = \int_{\U(d)} U^{\otimes N} \tilde{Q}_Y U^{\otimes N *} dU, 
  \end{equation}
  but this implies that $\tilde{Q}_Y$ must be of the form $\tilde{q}_Y \otimes \Bbb{1}$ with $q_Y \in
  \scr{B}(\scr{H}_Y)$. Hence (\ref{eq:50}) implies $q_{ZY} = 0$ for $Y \neq Z$, which proves the corollary.
\end{pf}

Since the estimation scheme $(\hat{F}_N)_{N \in \Bbb{N}}$ is asymptotically optimal, condition (\ref{eq:44})
looks at a first glance very natural. In contrast to permutation invariance and covariance, however, we have
no proof that it does not ``harm'' the rate function. In other words the crucial question is: Given a
covariant and permutation invariant estimation scheme $(E_N)_{N \in \Bbb{N}}$ satisfying LDP with rate function
$I$, does there exist a scheme $(\tilde{E}_N)_{N \in \Bbb{N}}$ which satisfies Equation (\ref{eq:44}) and the
LDP with a rate function $\tilde{I}$ such that $I \leq \tilde{I}$ holds?   A possible strategy towards a proof
might be to define $\tilde{E}_N$ by Equation (\ref{eq:31}) with $Q_Y = \int_\Sigma q_Y(dx)$ and the POV measures 
$q_Y$ which define $E_N$ according to Theorem \ref{prop:2}. The hard part (which we haven't solved up to now) 
is  of course to show that the rate function $\tilde{I}$ of such a scheme is at least as good as $I$. 

If we accept condition (\ref{eq:44}) nevertheless, the estimation scheme $(\hat{E}_N)_{N \in \Bbb{N}}$ arises
from Corollary \ref{kor:1} if we choose  
\begin{equation} \label{eq:52}
  Q_Y = \dim\scr{H}_Y \kb{\phi_Y},
\end{equation}
where $\phi_Y$ denotes the highest weight vector of the irreducible representation $\pi_Y$.
To see (heuristically) why this should be a good choice for the $Q_Y$, consider a nonsingular, diagonal
density matrix $\rho = e^h$ with $h=\diag(h_1,\ldots,h_d)$ and $h_1 \geq \cdots \geq h_d$. Since $\hat{E}_N$ projects to
$\hat{F}_N$ we know already that we get an exact estimate for the spectrum of $\rho$ in the limit $N \to \infty$. To get
a consistent scheme we need operators $Q_Y$ such that the quantities
\begin{equation} \label{eq:10}
  \tr \bigl( \pi_Y(U^*\rho U) Q_Y \bigr) \dim \scr{K}_Y = \tr \bigl( (U^*\rho U)^{\otimes N} (Q_Y \otimes \Bbb{1}) \bigr)
\end{equation}
(regarded as \emph{densities along the orbits} $\scr{S}_Y =  s^{-1}(Y/N)$) are more and more concentrated on
the density operators with the correct \emph{eigenvectors}, i.e. to $\rho_{Y/N}$. Since $Y \in \scr{Y}_d(N)$ is the
highest weight of the irreducible representation $\pi_Y$ and $\phi_Y$ its highest weight vector, the highest
eigenvalue of $\pi_Y(\rho)$ is given by $\exp(\sum_j Y_j h_j)$ and $\phi_Y \in \scr{H}_Y$ is the corresponding
eigenvector. All other eigenvalues grow with a lower exponential rate (or decay faster, depending on the 
chosen normalization). The matrix element $\langle\phi_Y, \pi_Y(\rho) \phi_Y\rangle$ dominates therefore all other eigenvalues in the
limit $N \to \infty$. Hence the density (\ref{eq:10}) has the desired behavior if we choose $Q_Y = \kb{\phi_Y}$. Note
that the reasoning just sketched indicate that for any consistent scheme of the form (\ref{eq:31}) the overlap
of the $Q_Y$ with $\kb{\phi_Y}$ should not decay too fast (at most polynomial). In the case of pure input state
we will make this reasoning more precise; cf. Section \ref{sec:pure-states}. 

%%%%%%%%%%%%%%%%%%%%%%%%%%%%%%%%%%%%%%%%%%%%%%%%%%
\subsection{Proof of Theorem \ref{thm:3}}
\label{sec:proof-theor-refthm:3}
%%%%%%%%%%%%%%%%%%%%%%%%%%%%%%%%%%%%%%%%%%%%%%%%%%

Our next task is to prove Theorem \ref{thm:3}, i.e. we have to show that the estimation scheme $\hat{E}_N$
defined in Equation (\ref{eq:8}) satisfies the LDP with rate function $\hat{I}$ given in (\ref{eq:11}). The
first step is to check that $\hat{I}$ is well defined.

\begin{lem}
  There is a (unique) function $\hat{I}$ on $\scr{S} \times \scr{S}$ which satisfies
  $\hat{I}(\rho,U\rho_xU^*) = \sum_{j=1}^d x_j \ln(x_j) - I_1(\rho,U,x)$ and
  \begin{equation} \label{eq:18}
    I_1(\rho,U,x) = \sum_{j=1}^d(x_j - x_{j+1}) \ln\bigl[\PM_j(U^*\rho U)\bigr]
  \end{equation}
  where we have set $x_{d+1}=0$. $\hat{I}$ is positive and $\hat{I}(\rho,\sigma) = 0$ implies $\sigma = \rho$.
\end{lem} 

\begin{pf}
  To prove that $\hat{I}$ is well defined we have to show that $U_1\rho_xU_1^* = U_2\rho_xU_2^*$ implies
  $I_1(\rho,U_1,x) = I_1(\rho,U_2,x)$. This is equivalent to  $[U,\rho_x] = 0$ $\Rightarrow$ $I_1(\rho,U,x) = I_1(\rho,\Bbb{1},x)$. To
  exploit the relation $[U,\rho_x] = 0$ let us introduce $k\leq d$ integers $1 = j_0 < j_1 <  
  \cdots < j_k = d+1$ such that $x_{j_\alpha} > x_{j_{\alpha+1}}$ and $x_j = x_{j_\alpha} > 0$ holds for $j_\alpha \leq j <
  j_{\alpha+1}$ and $\alpha < k$. Then we have 
  \begin{equation} \label{eq:15}
     I_1(\rho,U,x) = \sum_{\alpha=1}^k(x_{j_\alpha - 1} - x_{j_\alpha}) \ln\bigl[\PM_{j_\alpha - 1}(U^*\rho U)\bigr].
  \end{equation}
  On the other hand $[U,\rho_x] = 0$ implies that $U$ is block diagonal 
  \begin{equation}
      U = \diag(U_0, \ldots, U_{k-1})\  \text{with}\  U_\alpha \in  \U(d_\alpha),\ d_\alpha = j_{\alpha+1} - j_\alpha.
  \end{equation}
  Hence we have $\PM_{j_\alpha-1}(U^*\rho U) = \PM_{j_\alpha-1}(\rho)$ for all such $U$ and all $\alpha$ with $1 \leq \alpha \leq k$. Together
  with Equation   (\ref{eq:15}) this shows that $\hat{I}$ is well defined.  

  To prove positivity we have to show that $\inf_U \hat{I}(\rho,U\rho_xU^*) \geq 0$ holds for each $\rho$ and $x$. Hence we
  have to minimize $\hat{I}$ (for fixed $x$ and $\rho$) and since $x_j \geq x_{j+1}$ this implies that we have to
  maximize the minors of $U^*\rho U$. To this end let us denote the eigenvalues of $\rho$ and the upper left $j \times j$
  submatrix of $U^*\rho U$ by  $\lambda_1 \geq \lambda_2 \geq \cdots \geq \lambda_d$ respectively $\lambda^{(j)}_1 \geq \lambda_2^{(j)} \geq \cdots \geq \lambda_j^{(j)}$. The
  minors of $U^*\rho U$ then become $\PM_j(U^*\rho U) = \lambda^{(j)}_1 \cdots \lambda^{(j)}_j$. According to \cite[Thm
  4.3.15]{MR832183} the $\lambda^{(j)}_k$ satisfy the constraint $\lambda_k \geq \lambda^{(j)}_k$ for all $k=1,\ldots,j$, and this bound
  is (obviously) saturated if $U^*\rho U$ is diagonal in the preferred basis. Hence we get $\PM_j(U^*\rho U) \leq \lambda_1 \cdots
  \lambda_j$ and therefore
  \begin{equation}
    \hat{I}(\rho,U\rho_xU^*) \geq \sum_{j=1}^d x_j \ln(x_j) - \Sigma_{j=1}^d (x_j - x_{j+1}) \ln(\lambda_1 \cdots \lambda_j).
  \end{equation}
  Expanding the logarithms and reshuffling the second sum leads to
  \begin{equation}
    \hat{I}(\rho,U\rho_xU^*) \geq \sum_{j=1}^d x_j \bigl(\ln(x_j) - \ln(\lambda_j)\bigr),
  \end{equation}
  and equality holds iff $\rho$ and $\sigma = U\rho_xU^*$ are simultaneously diagonalizable. Since the left hand
  side of this inequality is a relative entropy of classical probability distributions, we see that $\hat{I}$
  is positive and $\hat{I}(\sigma) = 0$ holds iff $\sigma = \rho$. 
\end{pf} 

Now let us show that $(\hat{E}_N)_{N \in \Bbb{N}}$ satisfies the LDP with rate function $\hat{I}$. As in the
proof of Proposition \ref{prop:1} we will do this by proving the equivalent statement that $(\hat{E}_N)_{N \in
  \Bbb{N}}$ satisfies the Laplace principle with the same rate function, i.e.
\begin{equation}
  \lim_{N \to \infty} \frac{-1}{N} \ln \int_\scr{S} e^{-N f(\sigma)} \tr\bigl(\rho^{\otimes N} \hat{E}_N(d\sigma) \bigr) = \inf_{\sigma \in
    \scr{S}} \bigl( f(\sigma) + \hat{I}(\rho,\sigma) \bigr)
\end{equation}
should hold for all continuous functions $f$ on $\scr{S}$. If we insert the definition of $\hat{E}_N$, the
integral on the left hand side becomes
\begin{multline} \label{eq:17} 
  \int_\scr{S} e^{-N f(\sigma)} \tr\bigl(\rho^{\otimes N} \hat{E}_N(d\sigma) \bigr) = \\ \sum_{Y \in \scr{Y}_d(N)} \dim \scr{H}_Y
  \int_{\U(d)} e^{-N f(U\rho_{Y/N}U^*)} \tr\bigl((U^*\rho U)^{\otimes N} \kb{\phi_Y} \otimes \Bbb{1}_Y \bigr) dU,
\end{multline}
where $\Bbb{1}_Y$ denotes the unit operator on $\scr{K}_Y$. Now assume that $\rho$ is non-degenerate (i.e. $\rho \in
\GL(d,\Bbb{C})$) then we can rewrite the density in this integral to
\begin{align}
  \tr\bigl((U^*\rho U)^{\otimes N} \kb{\phi_Y} \otimes \Bbb{1}_Y\bigr) &= \tr \bigl( P_Y(U^*\rho U)^{\otimes N}P_Y\kb{\phi_Y}\otimes\Bbb{1}_Y \bigr) \\
  &=  \dim \scr{K}_Y \tr \bigl( \pi_Y(U^*\rho U) \kb{\phi_Y} \bigr) \\
  &=  \dim \scr{K}_Y \langle \phi_Y, \pi_Y(U^*\rho U) \phi_Y\rangle \label{eq:66}
\end{align}
where we have used in the second equation that $P_Y (U^*\rho U)^{\otimes N} P_Y = \pi_Y(U^*\rho U) \otimes \Bbb{1}_Y$ holds. The matrix
elements of $\pi_Y(U^*\rho U)$ with respect to the highest weight vector can be expressed as (\cite[{\S}
49]{Zhelo78} or \cite[Sect. IX.8]{Simon96}) 
\begin{equation} \label{eq:65}
  \langle\phi_Y, \pi_Y(U^*\rho U) \phi_Y\rangle = \prod_{k=1}^d \PM_k(U^*\rho U)^{Y_k - Y_{k+1}},
\end{equation}
where we have set $Y_{d+1}=0$. The right hand side of this equation makes sense even if the exponents are not
integer valued. We can rewrite therefore Equation (\ref{eq:17}) with the probability 
measure
\begin{equation} \label{eq:19}
  \int_\Sigma h(x) \nu_N(dx) = \frac{1}{d^N} \sum_{Y \in \scr{Y}_d(N)} h(\frac{Y}{N})
  \dim(\scr{H}_Y) \dim(\scr{K}_Y)
\end{equation}
to get
\begin{align}
  \int_\scr{S} e^{-N f(\sigma)}& \tr\bigl( \rho^{\otimes N} \hat{E}_N(d\sigma) \bigr) \\
  =& \int_\Sigma \int_{\U(d)} d^N e^{-N f(U\rho_xU^*)} \prod_{k=1}^d \PM_k(U^*\rho U)^{N(x_k-x_{k+1})} dU \nu_N(dx) \label{eq:67} \\
  =& \int_\Sigma \int_{\U(d)} \exp\bigl( -N \bigl[f(U \rho_x U^*) -\ln(d) - I_1(U,\rho,x)\bigr] \bigr) dU \nu_N(dx)  \label{eq:20}
\end{align}
where
\begin{equation}
  I_1(\rho,U,x) = \sum_{k=1}^d (x_k - x_{k+1}) \ln\bigl[\PM_k(U^*\rho U)\bigr] %\\
\end{equation}
is the function from Equation (\ref{eq:18}). Now we need the following Lemma

\begin{lem}
  The probability measures $\nu_N$ defined in Equation (\ref{eq:19}) satisfy the large deviation principle with
  rate function 
  \begin{equation}
    I_0(x) = \ln(d) + \sum_{j=1}^d x_j \ln(x_j).
  \end{equation}
\end{lem}

\begin{pf}
  This follows immediately from Theorem  \ref{thm:5} with $\rho = \frac{\Bbb{1}}{d}$ (cf. also
  \cite{Duffield90}).  
\end{pf}

Obviously the product measure $\nu_N(dx) \times dU$ satisfies the LDP with the same rate function. Moreover, the
function in the argument of the exponential in Equation (\ref{eq:20}) is continuous in $x$ and $U$. Hence we
can apply Varadhan's theorem to Equation (\ref{eq:20}) and get
\begin{align}
  \lim_{N \to \infty} \frac{-1}{N} \ln \int_\scr{S} & e^{-N f(\sigma)} \tr\bigl( \rho^{\otimes N} \hat{E}_N(d\sigma) \bigr) \\
  &= \inf_{x, U} ( f(U \rho_x U^*) - \ln(d) - I_1(U,\rho,x) + I_0(x))\\ 
  &= -\inf_{x, U} \left( f(U \rho_x U^*) + \sum_{j=1}^d x_j \ln(x_j) - I_1(U,\rho,x)\right),
\end{align}
which proves Theorem \ref{thm:3} for non-degenerate density matrices.

Now assume that $\rho$ is degenerate and has rank $r < d$. By continuity in $\rho$, Equations (\ref{eq:66}) and
(\ref{eq:65}) imply that 
\begin{equation} \label{eq:68}
    \tr\bigl((U^*\rho U)^{\otimes N} \kb{\phi_Y} \otimes \Bbb{1}_Y\bigr) = \dim \scr{K}_Y \prod_{k=1}^d \PM_k(U^*\rho U)^{Y_k - Y_{k+1}} 
\end{equation}
holds as in the non-degenerate case. The only difference is that the right hand side can vanish now, and it
vanishes in particular for all $Y$ with $Y_k > 0$ for $k > r$ (because all minors with $k > r$ vanish for any
$U$). Instead of (\ref{eq:67}) we therefore get 
\begin{multline} \label{eq:69}
  \int_\scr{S} e^{-N f(\sigma)} \tr\bigl( \rho^{\otimes N} \hat{E}_N(d\sigma) \bigr) \\  
  = \int_{\Sigma_r} \int_{\U(d)} r^N e^{-N f(U \rho_x U^*)} \prod_{k=1}^r \PM_k(U^*\rho U)^{N(x_k - x_{k+1})} dU \nu_{N,r}(dx)
\end{multline}
with
\begin{equation}
  \Sigma_r = \{ x \in \Sigma \, | \, x_k = 0\ \forall k> r \} 
\end{equation}
and 
\begin{equation}
  \int_{\Sigma_r} h(x) \nu_{N,r}(dx) = \frac{1}{d^N} \sum_{Y \in \scr{Y}_r(N)} h(\frac{Y}{N})
  \dim(\scr{H}_Y) \dim(\scr{K}_Y).
\end{equation}
Note that the difference between $\nu_N$ and $\nu_{N,r}$ is just the summation over all Young frames with $r$ rows
instead of $d$ rows. The right hand side of Equation (\ref{eq:68}) can still vanish because the unitary matrix
$U$ is a $d \times d$ matrix. Hence we can exclude
%\footnote{Note that $\scr{M}$ does depend on $\rho$ only  via its rank $r$.}
\begin{equation}
  \scr{M} = \{ U \in \U(d) \, | \, \PM_r(U^* \rho U) = 0\}
\end{equation}
from the domain of integration without changing the value of the integral in (\ref{eq:69}). Hence we get 
\begin{multline}
  \int_\scr{S} e^{-N f(\sigma)} \tr\bigl( \rho^{\otimes N} \hat{E}_N(d\sigma) \bigr) \\
  = \int_{\Sigma_r} \int_{\U(d) \setminus \scr{M}} \exp\bigl( -N \bigl[f(U \rho_x U^*) -\ln(r) - I_1(U,\rho,x)\bigr] \bigr) dU \nu_{N,r}(dx).
\end{multline}
The domain $\Sigma_r \times (\U(d) \setminus \scr{M})$ is open in $\Sigma_r \times \U(d)$ and $I_1$ is continuous on it. Hence we can
apply Varadhan's Theorem and proceed as in the non-degenerate case.

%%%%%%%%%%%%%%%%%%%%%%%%%%%%%%%%%%%%%%%%%%%%%%%%%%%%%%%%%%%%%%%%%%%%%%%%%%%%%%%%%%%%%%%%%%%%%%%%%%%%
\section{Upper bounds}
\label{sec:upper-bounds}
%%%%%%%%%%%%%%%%%%%%%%%%%%%%%%%%%%%%%%%%%%%%%%%%%%%%%%%%%%%%%%%%%%%%%%%%%%%%%%%%%%%%%%%%%%%%%%%%%%%%

In this section we will provide a detailed discussion of general upper bounds on admissible rate
functions. This includes in particular the proofs of Theorems \ref{thm:5} and \ref{thm:1}.

%%%%%%%%%%%%%%%%%%%%%%%%%%%%%%%%%%%%%%%%%%%%%%%%%%
\subsection{Hypothesis testing}
\label{sec:hypothesis-testing}
%%%%%%%%%%%%%%%%%%%%%%%%%%%%%%%%%%%%%%%%%%%%%%%%%%

Let us start with a very brief review of some material from quantum hypothesis testing (for a detailed
discussion cf. \cite{Helstrom,HolBook,HayashiBook}), because it can be used to derive related results for
estimation schemes. As in state estimation the task of hypothesis testing is to determine a state from
measurements on $N$ systems. In hypothesis testing, however, we know a priori that only a finite number of
different states can occur. For our  purposes it is sufficient to distinguish only between two states $\rho_0,\rho_1
\in \scr{S}$. This can be done by an observable of the $N$-fold system with values in the set $\{0,1\}$, where we
conclude from the outcome $j \in \{0,1\}$ that the initial preparation was done according to
$\rho_j$. Mathematically such an observable is given by a positive operator $A_N \in \scr{B}(\scr{H}^{\otimes N})$ with
$A_N \leq \Bbb{1}$ and $\tr(\rho_j^{\otimes N}A_N)$ is the probability to get the result $0$ during a measurement on $N$
systems in the joint state $\rho_j^{\otimes N}$. Hence the two quantities  
\begin{equation}
  \alpha_N(A_N) = \tr \bigl(\rho_0^{\otimes N}(\Bbb{1} - A_N)\bigr),\quad \beta_N(A_N) = \tr (\rho_1^{\otimes N}A_N)
\end{equation}
are error probabilities. More precisely $\alpha_N(A_N)$ is the probability to detect $\rho_1$ although the initial
preparation was given by $\rho_0^{\otimes N}$ (error of the first kind) and $\beta_N(A_N)$ is the probability for the converse
situation (error of the second kind). Ideally we would like to have a test which minimizes $\alpha_N$ \emph{and}
$\beta_N$. This is however impossible because we can always reduce one quantity at the expense of the other. A
possible solution of this problem is to make $\beta_N(A_N)$ as small as possible under the constraint that
$\alpha_N(A_N)$ remains bounded by some $\epsilon > 0$. The corresponding minimal (second kind) error probability is
therefore 
\begin{equation}
  \beta_N^*(\epsilon) = \inf \{ \beta_N(A_N) \, | \, A_N \in \scr{B}(\scr{H}^{\otimes N}),\ 0 \leq A_N \leq \Bbb{1},\ \alpha_N(A_N) \leq \epsilon\}. 
\end{equation}
Stein's Lemma describes the behavior of $\beta_N^*(\epsilon)$ in the limit $N \to \infty$; the quantum version is shown in
\cite{HP91,OgaNaga00}.  

\begin{thm}[Quantum Stein's Lemma] \label{thm:4}
  For any $0 < \epsilon < 1$ the equality
  \begin{equation}
    \lim_{N \to \infty} \frac{1}{N} \ln \beta_N^*(\epsilon) = - S(\rho_1,\rho_0)
  \end{equation}
  holds.
\end{thm}

%%%%%%%%%%%%%%%%%%%%%%%%%%%%%%%%%%%%%%%%%%%%%%%%%%
\subsection{State estimation} 
\label{sec:proof-theor-refthm:1}
%%%%%%%%%%%%%%%%%%%%%%%%%%%%%%%%%%%%%%%%%%%%%%%%%%

Let us consider now a (full) estimation scheme $(E_N)_{N \in \Bbb{N}}$. One possibility to distinguish between
two states $\rho$ and $\sigma$ is to choose a neighborhood $\Delta \in \goth{B}(\scr{S})$ of $\sigma$ with $\rho \not\in \Delta$ and to use
the tests $A_N = E_N(\Delta)$. If $(E_N)_{N \in \Bbb{N}}$ is consistent, the corresponding first kind error
probability $\alpha_N(A_N)$ vanishes in the limit $N \to \infty$ and we can apply Stein's Lemma to get a bound on
$\beta_N(A_N) = \tr\bigl(\rho^{\otimes N} E_N(\Delta)\bigr)$. Exploiting this idea more carefully leads to the following
theorem. 

\begin{thm} \label{thm:2}
  Consider a continuous map $p: \scr{S} \to X$ onto a locally compact, separable metric space $X$. The optimal
  rate function $\scr{I}_p$ defined in Equation (\ref{eq:22}) satisfies the inequality 
  \begin{equation} \label{eq:26}
    \scr{I}_p(\rho,x) \leq \inf_{\sigma \in p^{-1}(x)} S(\rho,\sigma)\quad \forall \rho \in \scr{S}\ \forall x \in X,
  \end{equation}
  where $S$ denotes the quantum relative entropy.
\end{thm}

\begin{pf}
  For each pair $\rho_0, \rho_1$ of density operators with $p(\rho_0) \neq p(\rho_1)$ we can find a sequence of tests
  $(A_N)_{N \in \Bbb{N}}$ by $A_N = E_N(\Delta)$ with an appropriate Borel set $\Delta \subset X$. If  $\Delta \in \goth{B}(X)$ is a
  neighborhood of $p(\rho_0)$, consistency of $(E_N)_{N \in \Bbb{N}}$ implies that for all $\epsilon >0$ there is an
  $N_\epsilon \in \Bbb{N}$ such that
  \begin{equation}
    \alpha_N(A_N) = 1 - \tr\bigl(E_N(\Delta) \rho_0^{\otimes N}\bigr) < \epsilon 
  \end{equation}
  holds for all $N > N_\epsilon$. Hence Stein's Lemma implies
  \begin{equation} \label{eq:13}
    \limsup_{N \to \infty} \frac{-1}{N} \ln \beta_N(A_N) = \limsup_{N \to \infty} \frac{-1}{N} \ln \tr \bigl( \rho_1^{\otimes N}
    E_N(\Delta)\bigr) \leq S(\rho_1,\rho_0). 
  \end{equation}
  Now assume that the rate function $I$ satisfies $I(\rho_1,x_0) > S(\rho_1,\rho_0)$ for some $\rho_0, \rho_1$ with $p(\rho_0) =
  x_0$ and $p(\rho_1) \neq x_0$. Since $I(\rho_1,\,\cdot\,)$ is lower semi-continuous we find a closed neighborhood $\Delta$ of
  $x_0$ such that   
  \begin{equation} \label{eq:12}
    I(\rho_1,x) \geq S(\rho_1,\rho_0) + \delta \quad \forall x \in \Delta  
  \end{equation}
  holds for an appropriate $\delta > 0$. Hence the large deviation upper bound (\ref{eq:2}) implies
  \begin{align}
    \limsup_{N \to \infty} \frac{1}{N} \ln \tr\bigl( \rho_1^{\otimes N} E_N(\Delta)\bigr) &\leq - \inf_{x \in \Delta} I(\rho_1,x)\\
    \liminf_{N \to \infty} \frac{-1}{N} \ln \tr\bigl( \rho_1^{\otimes N} E_N(\Delta)\bigr) &\geq \inf_{x \in \Delta} I(\rho_1,x)  \geq S(\rho_1,\rho_0) +
    \delta. \label{eq:14}
  \end{align}
  in contradiction to Equation (\ref{eq:13}). Hence $I(\rho_1,x_0) \leq S(\rho_0,\rho_1)$ for all $\rho_0$ with $p(\rho_0) =
  x_0$, which concludes the proof.
\end{pf}

\noindent \emph{Proof of Theorem \ref{thm:1}.} If we apply this theorem to full estimation schemes
(i.e. $X=\scr{S}$ and $p=\Id$) we get $I(\rho,\sigma) \leq S(\rho,\sigma)$ $\forall \rho,\sigma \in \scr{S}$ and Theorem \ref{thm:1} follows as a
simple corollary.\hfill $\Box$ \vspace{1em} 

\noindent \emph{Proof of Theorem \ref{thm:5}.} For a spectral estimation schemes with rate function $I$
Theorem \ref{thm:2} implies that $I(\rho,x) \leq \inf_{s(\sigma)=x}S(\rho,\sigma)$ holds. But the infimum on the right hand side
is achieved if $\sigma$ and $\rho$ commute and the eigenvalues in a joint eigenbasis are given in the same order. In
this case we have   
\begin{equation}
  S(\rho,\sigma) = \sum_{j=1}^d x_j\left(\ln x_j - \ln r_j\right) = S(x,r)
\end{equation}
where $s(\sigma) = x = (x_1,\ldots,x_d)$ and $s(\rho) = r = (r_1,\ldots,r_d)$ denote the ordered spectra of $\sigma$ and $\rho$ and
$S(r,x)$ is the \emph{classical} relative entropy of the probability vectors $r$ and $x$. Hence for spectral
estimation the upper bound (\ref{eq:26}) becomes
\begin{equation}
  I(\rho,x) \leq S\bigl(s(\rho),x\bigr)\quad \forall \rho \in \scr{S}\ \forall x \in \Sigma.
\end{equation}
But from \cite{KWEst} we know already that the scheme $(\hat{F}_N)_{N \in \Bbb{N}}$ defined in (\ref{eq:4})
saturates this bound; hence $(\hat{F}_N)_{N \in \Bbb{N}}$ is asymptotically optimal as stated in Theorem
\ref{thm:5}.\hfill $\Box$ \vspace{1em} 

If we are looking in particular at full estimation, the method used in the proof of Theorem \ref{thm:2} can be
improved significantly. The following lemma, which expresses the rate function explicitly as a limit over a
sequence of operators, is of great use in the next subsection. 

\begin{lem} \label{lem:1}
  Consider a full estimation scheme $(E_N)_{N \in \Bbb{N}}$ satisfying the LDP with rate function $I: \scr{S} \times
  \scr{S} \to [0,\infty]$ and two states $\rho,\sigma \in \scr{S}$. There is a sequence $(\Delta_N)_{N \in \Bbb{N}}$ of Borel sets
  $\Delta_N \subset \scr{S}$ satisfying 
  \begin{gather}
    \lim_{N \to \infty}\tr\bigl(\sigma^{\otimes N} E_N(\Delta_N) \bigr) = 1 \label{eq:25} \\
    \lim_{N \to \infty} \frac{-1}{N} \ln \tr\bigl(\rho^{\otimes N} E_N(\Delta_N)\bigr) =  I(\rho,\sigma) \label{eq:29}
  \end{gather}
  and
  \begin{equation} \label{eq:28}
    U \Delta_N U^* = \Delta_N\quad \forall U \in \U(d)\ \text{with}\ [U,\sigma] = 0
  \end{equation}
\end{lem}

\begin{pf}
  For each $k \in \Bbb{N}$ consider the set 
  \begin{equation}
    \tilde{\Delta}_k = \{ \omega \in \scr{S} \, | \, \|\sigma - \omega\|_1 \leq k^{-1} \} \subset \scr{S},
  \end{equation}
  which obviously has the symmetry property (\ref{eq:28}). Since the scheme $(E_N)_{N \in \Bbb{N}}$ is
  consistent (since $(E_N)_{N \in \Bbb{N}}$ satisfies the LDP this follows directly from Definition \ref{def:4})
  we have for each $k \in \Bbb{N}$ an index $N_k' \in \Bbb{N}$ such that 
  \begin{equation} \label{eq:45}
   \tr\bigl(\sigma^{\otimes N} E_N(\tilde{\Delta}_k) \bigr) \geq 1 - \frac{1}{k}
  \end{equation}
  holds for all $N \geq N_k'$. In addition we get for each $k \in \Bbb{N}$
    \begin{equation}
    \lim_{N \to \infty} \frac{-1}{N} \ln \tr\bigl(\rho^{\otimes N} E_N(\tilde{\Delta}_k)\bigr) = \inf_{\omega \in \tilde{\Delta}_k} I(\rho,\omega)
  \end{equation}
  by combining the large deviation upper and lower bounds. Hence for each $k \in \Bbb{N}$ there is an $N_k'' \in
  \Bbb{N}$ with
  \begin{equation} \label{eq:46}
        \left| \frac{-1}{N} \ln \tr\bigl(\rho^{\otimes N} E_N(\tilde{\Delta}_k)\bigr) - \inf_{\omega \in \tilde{\Delta}_k} I(\rho,\omega) \right| <
        \frac{1}{k} 
  \end{equation}
  for all $N \geq N_k''$. Now let us recursively define a strictly increasing sequence $(N_k)_{k \in \Bbb{N}}$ of
  integers by $N_1 = 1$ and $N_k = \max \{ N_k', N_k'', N_{k-1}+1\}$, and set 
  \begin{equation}
    \Delta_N = \tilde{\Delta}_k\ \text{for}\  N_k  \leq N < N_{k+1}.
  \end{equation}
  For each $N \geq N_k$ we therefore have an integer $l \geq k$ with $N_l \leq N < N_{l+1}$ and $\Delta_N =
  \tilde{\Delta}_l$. Since $N_l \leq N$ implies in particular $N \geq N_l'$ we have due to (\ref{eq:45})
  \begin{equation}
   \tr\bigl(\sigma^{\otimes N} E_N(\Delta_N) \bigr) = \tr\bigl(\sigma^{\otimes N} E_N(\tilde{\Delta}_l) \bigr) \geq 1 - \frac{1}{l} \geq 1 - \frac{1}{k}.
  \end{equation}
  and this implies Equation (\ref{eq:25}). Similarly we have $N \geq N_l \geq N_l''$ and therefore with
  (\ref{eq:46}) 
  \begin{multline} \label{eq:47}
    \left| \frac{-1}{N} \ln \tr\bigl(\rho^{\otimes N} E_N(\Delta_N)\bigr) - \inf_{\omega \in \Delta_N} I(\rho,\omega) \right| = \\
    \left| \frac{-1}{N} \ln \tr\bigl(\rho^{\otimes N} E_N(\tilde{\Delta}_l)\bigr) - \inf_{\omega \in \tilde{\Delta}_l} I(\rho,\omega) \right| <
    \frac{1}{l} \leq \frac{1}{k}.
  \end{multline}

  Now note that the sequence $(\Delta_N)_{N \in \Bbb{N}}$ forms a neighborhood base at $\sigma \in \scr{S}$, more precisely
  \begin{equation}
     \Delta_{N+1} \subset \Delta_N \ \forall N \in \Bbb{N}\ \text{and}\ \bigcap_{N=1}^\infty \Delta_N = \{\sigma\}.
  \end{equation}
  Lower semi-continuity of $I_\rho(\,\cdot\,) = I(\rho,\,\cdot\,)$ implies in addition that 
  \begin{equation}
    U_k = I_\rho^{-1}\bigl((I_\rho(\sigma) - k^{-1},\infty]\bigr)
  \end{equation}
  is for each $k \in \Bbb{N}$ an open neighborhood of $\sigma$. Hence we have a $M_k \in \Bbb{N}$ such
  that $M \geq M_k$ implies $\Delta_M \subset U_k$ and therefore  
  \begin{equation} \label{eq:24}
    I(\rho,\sigma) \geq \inf_{\omega \in \Delta_M} I(\rho,\omega) \geq I(\rho,\sigma) - \frac{1}{k}\quad \forall M \geq M_k.
  \end{equation}
  Now assume that $N \geq \max \{ N_k, M_k\}$ then we get with Equation (\ref{eq:47})
  \begin{multline}
    \left| \frac{-1}{N} \ln \tr\bigl(\rho^{\otimes N} E_N(\Delta_N)\bigr) - I(\rho,\sigma) \right| \leq \\
    \left| \frac{-1}{N} \ln \tr\bigl(\rho^{\otimes N} E_N(\Delta_N)\bigr) - \inf_{\omega \in \Delta_N} I(\rho,\omega) \right| + \\
    \left|\inf_{\omega \in \Delta_N} I(\rho,\omega) - I(\rho,\sigma)\right| \leq \frac{2}{k}
  \end{multline}
  and this implies Equation (\ref{eq:29}), which concludes the proof.
\end{pf}

%%%%%%%%%%%%%%%%%%%%%%%%%%%%%%%%%%%%%%%%%%%%%%%%%%
\subsection{Pure states}
\label{sec:pure-states}
%%%%%%%%%%%%%%%%%%%%%%%%%%%%%%%%%%%%%%%%%%%%%%%%%%

The main purpose of this section is to provide a proof of Equation (\ref{eq:36}), where we have claimed that 
$\hat{I}$ and $\scr{I}^c_{\Id}$ coincide for pure input states. This is basically quite simple. We will take,
however, a small detour which allows us to have a closer look beyond the covariant case (Subsection
\ref{sec:beyond-covariance}).

Let us consider first a pure state $\rho$ and a mixed state $\sigma$. From Equation (\ref{eq:11}) we see immediately
that this implies $\hat{I}(\rho,\sigma) = \infty$. Since $\hat{I}$ is a lower bound on all $\scr{I}_{\Id}^\#$ we
get
\begin{equation}
   \scr{I}_{\Id}(\rho,\sigma) = \scr{I}_{\Id}^0(\rho,\sigma) = \scr{I}_{\Id}^c(\rho,\sigma) = \hat{I}(\rho,\sigma) = \infty\quad \forall \rho\ \text{pure}\
    \sigma\ \text{mixed.}
\end{equation}
Hence only the case where $\rho$ and $\sigma$ are \emph{both pure} needs to be discussed. For the rest of this section
we will assume (unless something different is explicitly stated) therefore that   
\begin{equation} \label{eq:30}
  \rho = \kb{\phi},\ \sigma = \kb{\psi}\ \text{with}\ \phi, \psi \in \scr{H},\ \|\phi\|=\|\psi\|=1
\end{equation}
holds. The rate function $\hat{I}$ then has the following simple structure:
\begin{equation}
  \hat{I}(\rho,\sigma) = -\ln \tr(\rho\sigma) = - \ln \left(| \langle\phi,\psi\rangle |^2\right).
\end{equation}
Now we need the following lemma which shows that we can assume without loss of generality that the operators
$E_N(\Delta_N)$ from Lemma \ref{lem:1} are rank one projectors.

\begin{lem} \label{lem:2}
  Consider an admissible rate function $I \in \scr{E}(\Id)$ and two pure states $\rho = \kb{\phi}$,$\quad \sigma = \kb{\psi}$.
  There is a sequence $(\Psi_N)_{N \in \Bbb{N}}$ of normalized vectors $\Psi_N \in \scr{H}^{\otimes N}_+$ (the symmetric
  subspace of $\scr{H}^{\otimes N}$) such that 
  \begin{equation} \label{eq:56}
      \liminf_{N \to \infty} \frac{-1}{N} \ln  \left(|\langle\phi^{\otimes N}, \Psi_N\rangle|^2\right) \geq I(\rho,\sigma)
  \end{equation}
  and
  \begin{equation} \label{eq:55}
    \lim_{N \to \infty} |\langle\Psi_N,\psi^{\otimes N}\rangle|^2 = 1
  \end{equation}
  holds. If $I$ is covariant we can choose $\Psi_N = \psi^{\otimes N}$. 
\end{lem}

\begin{pf}
  Consider a full estimation scheme $(E_N)_{N \in \Bbb{N}}$ satisfying the LDP with rate function $I$ and the
  sequence $(\Delta_N)_{N \in \Bbb{N}}$ of Borel sets $\Delta_N \subset \scr{S}$ from Lemma \ref{lem:1}. Since only the overlap
  of $E_N(\Delta_N)$ with $\phi^{\otimes N}$ and $\psi^{\otimes N}$ are of interest, we can assume without loss of generality that
  $E_N(\Delta_N)$ is supported by the symmetric tensor product $\scr{H}^{\otimes N}_+$. Now choose a $0 < \lambda < 1$ and
  denote the spectral projector of $E_N(\Delta_N)$ belonging to the interval $[1-\lambda,1]$ by $P_{N,\lambda}$. Obviously we
  have due to $E_N(\Delta_N) \leq \Bbb{1}$   
  \begin{align}
   \langle\psi^{\otimes N}, E_N(\Delta_N) \psi^{\otimes N}\rangle &\leq \langle\psi^{\otimes N}, P_{N,\lambda} \psi^{\otimes N}\rangle \notag \\ 
   & \phantom{\leq \langle\psi^{\otimes N},}+ (1-\lambda) \langle\psi^{\otimes N},  (\Bbb{1} - P_{N,\lambda}) \psi^{\otimes N}\rangle \\[1em]
   &= (1-\lambda) + \lambda \langle\psi^{\otimes N}, P_{N,\lambda} \psi^{\otimes N}\rangle.
  \end{align}
  Equation (\ref{eq:25}) therefore implies
  \begin{equation}
    \lim_{N \to \infty} \langle\psi^{\otimes N}, P_{N,\lambda} \psi^{\otimes N}\rangle = 1
  \end{equation}
  Hence for each $0  < \delta < 1$ there is an $N_\delta \in \Bbb{N}$ such that 
  \begin{equation} \label{eq:54}
    \langle\psi^{\otimes N}, P_{N,\lambda} \psi^{\otimes N}\rangle \geq 1 -  \delta
  \end{equation}
  holds for all $N \geq N_\delta$. Now we define for $N$ with $P_{N,\lambda} \psi^{\otimes N} \neq 0$ (which due to Equation
  (\ref{eq:54}) is true if $N$ is large enough) 
  \begin{equation} \label{eq:49}
    \Psi_N = \frac{P_{N,\lambda}\psi^{\otimes N}}{\|P_{N,\lambda}\psi^{\otimes N}\|}
  \end{equation}
  and $\Psi_N$ arbitrary for all other $N$. Equation (\ref{eq:54}) implies immediately (\ref{eq:55}). The bound 
  (\ref{eq:56}) follows from
  \begin{equation}
    \langle\phi^{\otimes N}, E_N(\Delta_N) \phi^{\otimes N}\rangle \geq (1-\lambda) \langle\phi^{\otimes N}, P_{N,\lambda} \phi^{\otimes N}\rangle,
  \end{equation}
  which in turn implies
  \begin{align}
    I(\rho,\sigma) &= \lim_{N \to \infty} \frac{-1}{N} \ln \langle\phi^{\otimes N}, E_N(\Delta_N) \phi^{\otimes N}\rangle \\
    & \leq \liminf_{N \to \infty} \frac{-1}{N} \ln \langle\phi^{\otimes N}, (1-\lambda) P_{N,\lambda} \phi^{\otimes N}\rangle \\
    & = \liminf_{N \to \infty} \frac{-1}{N} \ln \langle\phi^{\otimes N}, P_{N,\lambda} \phi^{\otimes N}\rangle  \\
    & \leq \liminf_{N \to \infty} \frac{-1}{N} \ln \left(|\langle\phi^{\otimes N}, \Psi_N\rangle|^2\right)
  \end{align}
  where we have used in the last equation that $P_{N,\lambda} \Psi_N = \Psi_N$ and  therefore $P_{N,\lambda} \geq \kb{\Psi_N}$ holds
  if $N$ is large enough. 

  Now assume that $I$ is covariant. This implies by definition that we can choose $(E_N)_{N \in \Bbb{N}}$ to be
  covariant as well and we get according to Equation (\ref{eq:28}) 
  \begin{equation}
    U^{\otimes N} E_N(\Delta_N) U^{\otimes N *} = E_N(\Delta_N)\quad \forall U \in \U(d)\ \text{with}\ [U,\sigma] = 0. 
  \end{equation}
  Since $P_{N,\lambda}$ is a spectral projector of $E_N(\Delta_N)$ we get $U^{\otimes N} P_{N,\lambda} U^{\otimes N *} = P_{N,\lambda}$ for the
  same set of $U$ and since $\sigma = \kb{\psi}$ this implies
  \begin{equation}
    U^{\otimes N} P_{N,\lambda} \psi^{\otimes N} = P_{N,\lambda} U^{\otimes N} \psi^{\otimes N} = P_{N,\lambda} \psi^{\otimes N}\ \text{hence}\ U^{\otimes N} \Psi_N = \Psi_N
  \end{equation}
  for all $U$ with $U\psi = \psi$ and all $\Psi_N$ from Equation (\ref{eq:49}). It is easy to see that $\Psi_N = \psi^{\otimes N}$
  is the only vector in $\scr{H}^{\otimes N}_+$ with this property.
\end{pf}

With this lemma it is now very easy to determine $\scr{I}_{\Id}^c(\rho,\,\cdot\,)$ for pure input states $\rho$. As
already stated in Section \ref{sec:optim-rate-funct} we get (cf. in this context the analysis of covariant
pure state estimation in \cite{Hayashi98}) 

\begin{prop} \label{thm:10}
  For each pure state $\rho$ and all $\sigma \in \scr{S}$ the equality 
  \begin{equation} %\label{eq:36}
    \scr{I}^c_{\Id}(\rho,\sigma) = \hat{I}(\rho,\sigma) =  
    \begin{cases}
      \infty & \text{if $\sigma$ is mixed}\\
      - \ln \tr(\rho\sigma) & \text{if $\sigma$ is pure}
    \end{cases}
  \end{equation}
  holds.
\end{prop}

\begin{pf}
  Since $\hat{I}$ is covariant we have $\scr{I}^c_{\Id}(\rho,\sigma) \geq \hat{I}(\rho,\sigma)$ for all $\rho,\sigma \in \scr{S}$. If $\rho$
  is pure and $\sigma$ is mixed we have $\hat{I}(\rho,\sigma) = \infty$ and therefore $\scr{I}^c(\rho,\sigma) = \hat{I}(\rho,\sigma)$. If both
  states are pure we get from Lemma \ref{lem:2}
  \begin{equation}
    \scr{I}^c(\rho,\sigma) \leq \lim_{N \to \infty} \frac{-1}{N} \ln |\langle\phi^{\otimes N}, \psi^{\otimes N}\rangle|^2 = - \ln \tr(\rho\sigma) = \hat{I}(\rho,\sigma)
  \end{equation}
  which concludes the proof. 
\end{pf}

Together with the arguments from Section \ref{sec:an-explicit-scheme} this result supports our conjecture from
Section \ref{sec:optim-rate-funct} that $\scr{I}_{\Id}^c$ and $\hat{I}$ coincide also for mixed input states.

%%%%%%%%%%%%%%%%%%%%%%%%%%%%%%%%%%%%%%%%%%%%%%%%%%
\subsection{Beyond covariance}
\label{sec:beyond-covariance}
%%%%%%%%%%%%%%%%%%%%%%%%%%%%%%%%%%%%%%%%%%%%%%%%%%

If we look at Equation (\ref{eq:55}) and compare it with the reasoning in the last proof we might think that 
covariance is not really needed here, because $\Psi_N$ converges to $\psi^{\otimes N}$ in the limit $N \to \infty$ even without
further assumptions on $I$. This impression, however, is wrong, because the vectors $\psi^{\otimes N}$ and $\phi^{\otimes N}$
become more and more orthogonal as $N$ increases and therefore the part of $\Psi_N$ which is orthogonal to $\psi^{\otimes
  N}$ can play a crucial role (although it vanishes in the limit $N \to \infty$). The relation of the optimal
rate functions $\scr{I}_{\Id}$ and $\scr{I}_{\Id}^0$ to $\hat{I}$ and relative entropy $S$ needs therefore
more discussion. Although we are not yet able to give complete results we will collect in the following some
(informal) arguments which supports the two conjectures $\scr{I}_{\Id} = S$ and $\scr{I}_{\Id}^0 = \hat{I}$
from the end of Section \ref{sec:optim-rate-funct}. 

As in the last section we will consider only pure states, i.e. we will evaluate a rate function $I(\rho,\sigma)$ only
for $\rho = \kb{\phi}$ and $\sigma = \kb{\psi}$. In addition we will assume that $\scr{H}$ is two-dimensional (this can be
done without loss of generality, because we just have to replace $\scr{H}$ with the subspace generated by $\psi$ and
$\phi$). Hence we can set
\begin{equation} \label{eq:74}
  \psi = \ket{0}\ \text{and}\ \phi = \phi_{p,\alpha} = \sqrt{p} \ket{0} + e^{i\alpha} \sqrt{1-p} \ket{1}
\end{equation}
with $0 \leq p \leq 1$, $\alpha \in (-\pi,\pi]$ and an arbitrary but fixed basis $\ket{0}, \ket{1}$ of $\scr{H}$. In the number
basis $\ket{k,N} \in \scr{H}^{\otimes N}_+$, $k = 0,\ldots,N$ 
\begin{equation}
  \ket{k;N} = \binom{N}{k}^{-1/2} S_N \ket{0}^{\otimes (N-k)} \otimes \ket{1}^{\otimes k}
\end{equation}
(where $S_N$ is the projector to $\scr{H}^{\otimes N}_+$) the vectors $\Psi_N \in \scr{H}^{\otimes N}_+$ from Lemma \ref{lem:2}
can then be written as 
\begin{equation} \label{eq:75}
  \Psi_N = \sum_{k=0}^N f_{N,k} \ket{k;N} 
\end{equation}
and $\phi^{\otimes N}$ becomes
\begin{equation}
  \phi^{\otimes N} = \phi_{p,\alpha}^{\otimes N} = \sum_{k=0}^N \binom{N}{k}^{1/2} \sqrt{p}^{N-k} \sqrt{1-p}^k e^{ik\alpha} \ket{k,N}.
\end{equation}

Let us consider the conjecture $\scr{I}_{\Id} = S$ first. In the case of pure states this would imply that we 
can find for each pair of pure states $\sigma \neq \rho_0$ an admissible rate function $I$ with $I(\rho_0,\sigma) = \infty$. A
possible way to prove this could consist of two steps:
\begin{itemize}
\item 
  \emph{Step 1.} Find a sequence $(A_N)_{N \in \Bbb{N}}$ of operators such that 
  \begin{equation} \label{eq:72}
    \lim_{N \to \infty} \frac{-1}{N} \ln \tr(\rho_0^{\otimes N} A_N) = \infty,\quad \lim_{N \to \infty} \tr(\sigma^{\otimes N} A_N) = 1
  \end{equation}
  and 
  \begin{equation} \label{eq:73}
    \lim_{N \to \infty} \frac{-1}{N} \ln \tr(\rho^{\otimes N} A_N) = I^\sigma(\rho) > 0\quad \forall \rho \neq \sigma
  \end{equation}
  holds.
\item  
  \emph{Step 2.} Find a full estimation scheme $(E_N)_{N \in \Bbb{N}}$ and a sequence $(\Delta_N)_{N \in \Bbb{N}}$ of
  Borel sets $\Delta_N \subset \scr{S}$ shrinking to $\sigma$ such that $E_N(\Delta_N) = A_N$ holds for all $N \in \Bbb{N}$.  
\end{itemize}
To implement the second step we would need a converse of Lemma \ref{lem:1}, and such a result is
(unfortunately) not yet available. The problem here is not to construct some POV measures with $E_N(\Delta_N) =
A_N$, but to construct them such that the resulting scheme satisfies the LDP (which includes in particular 
consistency). It seems, however, that this is more a technical then a fundamental problem. 

The first step is much easier to perform\footnote{However, it is not sufficient to find a sequence of tests
  which saturates the bound from Stein's lemma, because Equation (\ref{eq:73}) would not necessarily hold in
  this case.}. Assume that $\rho_0 = \kb{\phi_{q,\beta}}$ holds with $\phi_{q,\beta}$ from
(\ref{eq:74}). Then we set $A_N = \kb{\Psi_N}$ and define $\Psi_N$ according to (\ref{eq:75}) by
\begin{equation}
  f_{N,0} = - \scr{N}_N \sqrt{N} \sqrt{1-q} e^{i\beta},\quad f_{N,1} = \scr{N}_N \sqrt{q}
\end{equation}
with the normalization 
\begin{equation}
  \scr{N}_N = \bigl(N (1-q) + q\bigr)^{-1/2}
\end{equation}
and $f_{N,k} = 0$ for all $k > 1$. Obviously we have 
\begin{equation}
  \langle\Psi_N,\phi^{\otimes N}_{q,\beta}\rangle = 0\ \text{and}\ \lim_{N \to \infty} f_{N,0} = 1
\end{equation}
which implies Equations (\ref{eq:72}). On the other hand we get $I^\sigma(\rho) = - \ln \tr (\rho \sigma)$ for each pure $\rho \neq
\rho_0$ and therefore Equation (\ref{eq:73}) holds as well. Hence there is strong evidence behind the conjecture
$\scr{I}_{\Id} = S$ from Section \ref{sec:optim-rate-funct} (at least for pure input states).   

The method used in the last paragraph can be easily generalized to construct a sequence of operators $(A_N)_{N
  \in \Bbb{N}}$ such that the function $I^\sigma$ from (\ref{eq:73}) becomes infinite at finitely many points or even
on a countable dense subset of the space $\scr{P}$ of pure states. This is, however, not sufficient to disprove the
conjecture $\scr{I}_{\Id}^0 = \hat{I}$, because in this case we would need $(A_N)_{N \in \Bbb{N}}$ such that
$I^\sigma$ becomes lower semi-continuous. $I^\sigma(\rho_0) > - \ln \tr (\rho_0\sigma)$ for one state $\rho_0$ implies for such an $I^\sigma$
that $I^\sigma(\rho) > - \ln(\rho\sigma)$ holds for all $\rho$ in a whole neighborhood of $\rho_0$ in $\scr{P}$. We will show
in the following why it is (at least) very difficult to find a sequence $(A_N)_{N \in \Bbb{N}}$ with this
special property. 

To this end consider $A_N = \kb{\Psi_N}$ with $\Psi_N$ from Lemma \ref{lem:2} and a fixed $0 < p < 1$ such that    
\begin{equation} \label{eq:76}
  \lim_{N \to \infty} \frac{-1}{N} \ln \left( |\langle\Psi_N, \phi^{\otimes N}_{p,\alpha}\rangle|^2\right) > - \ln \tr
  \left(|\langle\psi,\phi_{p,\alpha}\rangle|^2\right) = - \ln p  
\end{equation}
holds for all $\alpha$ with $-\pi < \alpha_- < \alpha < \alpha_+ < \pi$ for some bounds $\alpha_-,\alpha_+$. To rewrite this in a more convenient
way let us identify the interval $(-\pi,\pi]$ with the unit circle $S^1$ and consider the sequence $(F_N)_{\Bbb{N} \in
  \Bbb{N}}$, $F_N \in \Lz(S^1)$ with
\begin{equation}
  F_N = \|\tilde{F}_N\|^{-1} \tilde{F}_N,\quad \tilde{F}_N(\alpha) = \langle\Psi_N, \phi^{\otimes N}_{p,\alpha}\rangle.
\end{equation}
In the orthonormal basis $(e_k)_{k \in \Bbb{Z}}$, $e_k \in \Lz(S^1)$, $e_k(\alpha) = (2\pi)^{-1/2} \exp(i k \alpha)$ these
vectors become 
\begin{equation}
  \tilde{F}_N(\alpha) = \sum_{k = 0}^N \overline{f_{N,k}} \binom{N}{k}^{1/2} \sqrt{p}^{N-k} \sqrt{1-p}^k e^{ik\alpha},
\end{equation}
hence all $F_N$ are elements of the positive frequency subspace
\begin{equation}
  \Hz(S^1) = \SP \{ e_k \, | \, k \geq 0 \} \subset \Lz(S^1).
\end{equation}
In addition we can conclude immediately from Equation (\ref{eq:55}) and $\ket{0,N} = \psi^{\otimes N}$ the inequality
\begin{equation}
  \lim_{N \to \infty} \frac{-1}{N} \ln \left( \|\tilde{F}_N\|^2\right) \leq - \ln p.
\end{equation}
Hence to get (\ref{eq:76}) the functions $F_N$ have to converge pointwise and exponentially fast to $0$ on
the interval $(\alpha_-,\alpha_+)$. 
%Since all $F_N$ are continuous the convergence is even uniform on every compact subinterval. 
To find such a sequence is difficult due to the following lemma.

\begin{lem}
  A function $F \in \Hz(S^1)$ which vanishes on a non-empty subinterval $(\alpha_-,\alpha_+)$ of $S^1$ vanishes completely.
\end{lem}

The proof of this lemma uses the fact that each smooth element of $\Hz(S^1)$ is the boundary value of an
analytic function on the unit disc (cf. \cite{MR1645078} for details). For us it shows that the $F_N$ can not
vanish on $(\alpha_-,\alpha_+)$ because $\|F_N\|=1$ by construction. It is even impossible that the sequence $(F_N)_{N
  \in \Bbb{N}}$ converges (in norm) to a function $F \in \Lz(S^1)$, because this $F$ would satisfy again $\|F\|=1$,
$F \in \Hz(S^1)$ and $F(\alpha) = 0$ for all $\alpha \in (\alpha_-,\alpha_+)$. The only way out is to find a sequence which does not
converge for all $\alpha$. Such a series can be constructed if we allow infinitely fast oscillations in the limit
$N \to \infty$ (start with a sequence which converges in $\Lz(S^1)$ and shift its elements to the positive frequency
space). However, even then there are two additional requirements: 1. The vectors $\Psi_N$ (and therefore the
coefficients $f_{N,k}$) have to satisfy the constraints $\|\Psi_N\|=1$ and $\lim_{N \to \infty} |f_{N,0}| = 1$ and
2. $\lim_{N \to \infty} F_N(\alpha) =0$ must hold not only for all $\alpha \in (\alpha_-,\alpha_+)$, but also for all $p \in (p_-,p_+)$ for
some $0 < p_- < p_+ < 1$. We have not yet succeeded to construct a sequence $(\Psi_N)_{N \in \Bbb{N}}$ which satisfies
all these condition, but what we can say already at this point is the following: If there is a rate function
$I \in \scr{E}^0(\Id)$ with $I(\rho,\sigma) > \hat{I}(\rho,\sigma)$ for some $\rho,\sigma$, then the corresponding estimation scheme
must develop very irregular behavior with respect to relative phases and this indicates that a more detailed
analysis of phase estimation might solve our problem.

\begin{appendix}
  
%%%%%%%%%%%%%%%%%%%%%%%%%%%%%%%%%%%%%%%%%%%%%%%%%%%%%%%%%%%%%%%%%%%%%%%%%%%%%%%%%%%%%%%%%%%%%%%%%%%%
\section{Some material from large deviations theory}
\label{sec:some-material-from}
%%%%%%%%%%%%%%%%%%%%%%%%%%%%%%%%%%%%%%%%%%%%%%%%%%%%%%%%%%%%%%%%%%%%%%%%%%%%%%%%%%%%%%%%%%%%%%%%%%%%

The purpose of this appendix is to collect some material about large deviation theory which is used
throughout this paper. For a more detailed presentation we refer the reader to monographs like
\cite{Ellis85,DupEll,MR1739680}.  

\begin{defi} \label{def:1}
  A function $I: X \to [0,\infty]$ on a locally compact, separable, metric space $X$ is called a \emph{rate function}
  if 
  \begin{enumerate}
  \item 
    $I \not \equiv \infty$
  \item 
    $I$ is lower semi-continuous.
  \item 
    $I$ has compact level sets, i.e. $I^{-1}\bigl([-\infty,c]\bigr)$ is compact for all $c \in \Bbb{R}$.
  \end{enumerate}
\end{defi}

\begin{defi} \label{def:3}
  Let $(\mu_N)_{N \in \Bbb{N}}$, $N \in \Bbb{N}$ be a sequence of probability measures on the Borel subsets of a locally compact,
  separable metric space $X$ and $I: X \to [0,1]$ a rate function in the sense of Definition \ref{def:1}.  We
  say that $(\mu_N)_{N \in \Bbb{N}}$ satisfies the \emph{large deviation principle} with  rate function $I: X \to
  [0,\infty]$ if the following conditions hold:  
  \begin{enumerate}
  \item 
    For each closed subset $\Delta \subset \Sigma$ we have
    \begin{equation}  \label{eq:2}
      \limsup_{N \to \infty} \frac{1}{N} \ln \mu_N(\Delta) \leq - \inf_{x \in \Delta} I(x)
    \end{equation}
  \item 
    For each open subset $\Delta \subset \Sigma$ we have
    \begin{equation}
      \liminf_{N \to \infty} \frac{1}{N} \ln \mu_N(\Delta) \geq - \inf_{x \in \Delta} I(x)
    \end{equation}
  \end{enumerate}
\end{defi}

The most relevant consequence of this definition is the following theorem of Varadhan \cite{Varadhan66},
which describes the behavior of some expectation values in the limit $N \to \infty$:

\begin{thm}[Varadhan]  \label{thm:7}
  Consider a sequence $(\mu_N)_{N \in \Bbb{N}}$, $N \in \Bbb{N}$ of probability measures on $X$ satisfying the large
  deviation principle with rate function $I: X \to [0,\infty]$ and a continuous function $f: X \to \Bbb{R}$ which is
  bounded from below. Then the following equality holds:
  \begin{equation} \label{eq:16}
    \lim_{N \to \infty} \frac{1}{N} \ln \int_E e^{-N f(x)} \mu_N(dx) = - \inf_{x \in E}
    \bigl( f(x) + I(x) \bigr). 
  \end{equation}
\end{thm}

Varadhan's theorem has a converse: If we know that a sequence of measures $\mu_N$ satisfies Equation
(\ref{eq:16}) for all bounded continuous functions it can be shown that the $\mu_N$ satisfy the
large deviation principle as well. Following \cite{DupEll} we have:

\begin{defi} \label{def:5}
  Let $(\mu_N)_{N \in \Bbb{N}}$ be a sequence of measures on a locally compact, separable metric space $X$ and $I:
  X \to [0,\infty]$ a rate function. We say that $(\mu_N)_{N \in \Bbb{N}}$ satisfy the \emph{Laplace principle} with rate
  function $I$, if we have 
  \begin{equation}
    \lim_{N \to \infty} \frac{1}{N} \ln \int_E e^{-N f(x)} \mu_N(dx) = - \inf_{x \in E} \bigl( f(x) + I(x) \bigr). 
  \end{equation}
  for all bounded continuous functions $f: E \to \Bbb{R}$.
\end{defi} 

\begin{thm} \label{thm:8}
  The Laplace principle implies the large deviation principle with  the same rate function.  
\end{thm}

\end{appendix}

\section*{Acknowledgments}

I would like to thank R. D. Gill and R. F. Werner for many useful discussions, and M. Hayashi for comments on
an earlier version of this manuscript. Financial support by the European Union project ATESIT (contract
no. IST-2000-29681) is also greatfully acknowledged. 

\bibliographystyle{mk}
\bibliography{qinf}

\end{document}